\pdfoutput=1

\documentclass[aps,prd,amsmath,floats,floatfix,superscriptaddress,nofootinbib,showpacs]{revtex4-2}

\usepackage[T1]{fontenc}
\usepackage[utf8]{inputenc}
\usepackage{lmodern}
\usepackage{xfrac}
\usepackage{verbatim}
\usepackage{footmisc}

\usepackage[dvipsnames, usenames]{xcolor}
\definecolor{linkcolor}{rgb}{0.0,0.3,0.5}
\usepackage[hypertexnames=false, unicode, colorlinks=true, linkcolor=magenta,
citecolor=magenta, filecolor=magenta,urlcolor=magenta,
pdfusetitle]{hyperref}
\usepackage{url}
\usepackage[all]{hypcap}
\usepackage{graphicx}
\usepackage{subcaption}
\usepackage{xspace}
\usepackage{amsmath}
\usepackage{amssymb}
\usepackage{siunitx}
\DeclareSIUnit \parsec {pc}
\DeclareSIUnit \h {h}
\usepackage[normalem]{ulem} 
\usepackage{bm} 

\usepackage{microtype}

\usepackage[english]{babel}
\usepackage{blindtext}
\usepackage{natbib}
\usepackage{comment}

\graphicspath{%
  {figs/}%
}

\DeclareMathAlphabet{\mathpzc}{OT1}{pzc}{m}{it}

\renewcommand{\vec}[1] {\boldsymbol{#1}} 
\newcommand*{\vq} {\vec{q}}

\newcommand*{\vx} {\vec{x}}
\newcommand*{\vk} {\vec{k}}
\newcommand*{\vv} {\vec{v}}
\newcommand*{\vw} {\vec{\omega}}
\newcommand*{\p}  {\partial}
\newcommand*{\df}  {\delta}
\newcommand*{\tf}  {\theta}

\newcommand*{\non}  {\nonumber}

\newcommand{\ba}{\[\begin{aligned}}
\newcommand{\ea}{\end{aligned}\]}

\newcommand{\eq}[1]{\begin{align}#1\end{align}}
\newcommand{\eeq}[1]{\begin{equation}#1\end{equation}}

\begin{document}

\rightline{\scriptsize RBI-ThPhys-2023-11}
\title{
Optimizing the Evolution of Perturbations in the $\Lambda$CDM Universe
}

\newcommand\nickhomeone{\affiliation{Sub-department of Astrophysics, University of Oxford, Keble Road, Oxford OX1 3RH, UK}}
\newcommand\anthonyhomeone{\affiliation{Institute of Astronomy, Madingley Road, Cambridge, CB3 0HA, UK}}
\newcommand\anthonyhometwo{\affiliation{Kavli Institute for Cosmology Cambridge, Madingley Road, Cambridge, CB3 0HA, UK}}
\newcommand\anthonyhomethree{\affiliation{DAMTP, Centre for Mathematical Sciences, Wilberforce Road, Cambridge CB3 0WA, UK}}
\newcommand\zvonehomeone{\affiliation{Division of Theoretical Physics, Ru\dj er Bo\v{s}kovi\'c Institute, 10000 Zagreb, Croatia}}

\author{Nicholas Choustikov}
\email{nicholas.choustikov@physics.ox.ac.uk}
\nickhomeone
\anthonyhomeone

\author{Zvonimir Vlah}
\email{zvlah@irb.hr}
\zvonehomeone
\anthonyhometwo
\anthonyhomethree

\author{Anthony Challinor}
\email{a.d.challinor@ast.cam.ac.uk}
\anthonyhomeone
\anthonyhometwo
\anthonyhomethree

\hypersetup{pdfauthor={Choustikov et al.}}


\begin{abstract}
\vspace{.5cm}
\noindent 
Perturbation theory is a powerful tool for studying large-scale structure formation in the universe and calculating observables such as the power spectrum or bispectrum.
However, beyond linear order, typically this is done by assuming a simplification in the time-dependence of gravitational-coupling kernels between the matter and velocity fluctuations. Though the true dependencies are known for Lambda cold dark matter cosmologies, they are ignored due to the computational costs associated with considering them in full and, instead, are replaced by simpler dependencies valid for an Einstein--de-Sitter cosmology. Here we develop, implement and demonstrate the effectiveness of a new numerical method for finding the full dynamical evolution of these kernels to all perturbative orders based upon spectral methods using Chebyshev polynomials. This method is found to be orders of magnitude more efficient than direct numerical solvers while still producing highly accurate and reliable results. A code implementation of the Chebyshev spectral method is then presented and characterised. The code has been made publicly available alongside this paper. We expect our method to be of use for interpretation of upcoming galaxy clustering measurements.
\end{abstract}

\maketitle
\section{Introduction} 
\label{sec:intro}
\noindent 

Studying the history of large-scale structure (LSS) formation in our Universe is crucial to modern cosmology. The highly structured cosmic web of galaxy clusters, sheets, walls, filaments and voids present in the late-time Universe arises from evolution under gravity of small, primordial fluctuations in the density and velocity of matter, believed to have been sourced during a period of cosmic inflation. LSS therefore encodes information on the primordial fluctuations as well as the expansion history, geometry and matter content of the Universe, which affect the subsequent evolution of the perturbations.

Theoretical studies of LSS aim to predict the statistical properties of the clustering of the matter density (and associated velocity), for example, the power spectrum and higher-point correlation functions. These can be compared with the observed statistics of the clustering of galaxies since the galaxy over-density on large scales traces the matter over-density. Such comparisons are complicated by two issues, however. Firstly, galaxies need not necessarily follow the underlying matter distribution exactly, although on the largest cosmological scales the relation for galaxies is essentially linear with a constant of proportionality known as galaxy bias, while on smaller, mildly nonlinear scales, corrections can be treated perturbatively~\citep{Desjacques:2018bnm}. Secondly, we have the issue that each galaxy’s redshift depends not only on distance but also on its peculiar velocity via the Doppler effect~\citep{Kaiser:1992, Hamilton:1997}. Furthermore, these velocities are not random but instead correlate with the matter density field itself. This alters galaxy statistics by producing redshift-space distortions (RSDs), also introducing features such as the well-known ``Fingers of God''~\citep{Hamilton:1997}. These issues are compounded by the fact that we are now entering the era of high-precision cosmology with large galaxy surveys such as Euclid~\citep{Amendola:2016}, DESI~\citep{Aghamousa:2016}, Rubin~\citep{Abate:2012} and Roman~\citep{Spergel:2015}. Here, it becomes necessary to quantify these effects fully, lest they introduce greater uncertainties or issues of biased parameter estimation.

Given the small amplitude of the primordial fluctuations, perturbation theory is a very powerful tool for calculating the evolution of fluctuations in cosmological fields. For early-time observables, such as the cosmic microwave background (CMB), linear perturbation theory is an excellent description. Accurate predictions for the CMB power spectra can be calculated in this way (e.g.,~\cite{Peebles:1970ag, Bond:1984fp}) and their comparison with measurements of the CMB temperature and polarization anisotropies has been instrumental in establishing the standard model of cosmology, the Lambda cold dark matter ($\Lambda$CDM) model, and determining its parameters to high precision (e.g.,~\cite{Planck:2018vyg}). However, linear perturbation theory fails to describe the late-time Universe, where, particularly at smaller scales, complex non-linearities take hold and baryon physics begins to play a role~\citep{Chisari:2018prw}. Here, higher-order perturbation theory is crucial to building analytic models. The standard procedure in these perturbative approaches is to use the quasi-static Einstein--de-Sitter (EdS) approximation, henceforth referred to as the quasi-EdS approximation (qEDS). Here, a blend of models is used, with non-linear interaction kernels calculated in the dramatically simpler EdS (completely matter-dominated) universe. These are then combined with linear growth-rate results for the full $\Lambda$CDM dynamical Universe in order to extract observables \cite{Bernardeau:2001qr}. In essence, this approximation assumes that any non-linearities are constant in time, to then be scaled up to their present-day values.

The qEDS approach has been fairly successful thus far, yielding percent-level accuracy for all necessary two-point statistics close to the linear regime \citep{Fasiello+:2016}. Such models have already yielded concrete predictions, in particular for the one-loop power spectrum and tree-level bispectrum, and these have been used to extract relevant cosmological information from galaxy surveys \citep{DAmico:2019fhj, Ivanov:2019pdj, Chen:2021}. However, for future surveys, this accuracy will not be enough. Particularly in the context of RSDs, the full dynamics of $\Lambda$CDM must be taken into account. A large body of work has already explored the extent of these effects at the level of one-loop results \cite{Bernardeau:1993,Takahashi:2008,Fasiello:2016,Lewandowski:2016,Fujita:2020xtd,Donath:2020abv,Schmidt:2020,Rampf:2022tpg}, and recently to two-loop order\citep{Garny:2020,Fasiello:2022lff,Garny:2022fsh}. Unfortunately, accounting for the full $\Lambda$CDM dynamics introduces several issues, including that some of these features are slow to calculate using brute-force numerical methods (especially at higher perturbative order). This makes exploration of parameter space computationally challenging when interpreting survey data.
Therefore, this paper aims to optimize the process by which the full $\Lambda$CDM dynamics can be calculated at each perturbative order, using previous results for time-dependent coefficients as a starting point~\citep{Fasiello:2016, Fasiello:2022lff}.

{We tackle this problem with a novel numerical method, expanding the time-dependent interaction kernels in shifted Chebyshev polynomials~\citep{Karjanto:2020} as in the Chebyshev Spectral Method (CSM; see, e.g.,~\cite{Sezer:1996} for the explicit, matrix-based approach that we follow).
Doing so, we convert the process of calculating these dynamical coefficients into a linear-algebra problem that can readily be solved. We have implemented this method into a new Python library that we present below. 
The code is freely available at \url{https://github.com/Chousti/CSMethod.git},
with the hope that it can be integrated into likelihood analyses of forthcoming survey data, allowing
more accurate parameter estimates to be derived from these data.
While we focus on $\Lambda$CDM cosmologies here, it is expected that the method can be straightforwardly extended to
more generalised cosmologies, including features such as clustered quintessence~\citep{Sefusatti:2011cm, Fasiello:2019}.

This paper is arranged as follows. In Sec.~\ref{sec:LCDM_dynamics} we review certain theoretical results underpinning the problem. Section~\ref{sec:spectral} discusses the implementation of our spectral method with explanations of the code philosophy, and Sec.~\ref{sec:results} presents results and code tests. Finally, we discuss our results in Sec.\ref{sec:conclusion}. More extensive auxiliary functions are given in Appendix~\ref{sec:appendixI}.

All numerical results (unless stated otherwise) are calculated using the Planck best-fit $\Lambda$CDM cosmology, with parameters as follows: the present-day matter density parameter $\Omega_{m_0} = 0.315$; present-day dark-energy density parameter $\Omega_{\Lambda_0} = 0.685$, giving a flat universe; and Hubble constant $H_0 = \SI{67.74}{\kilo\metre\per\second\per\mega\parsec}$~\citep{Planck:2018vyg}.

\section{Dynamics in the $\Lambda$CDM Universe}
\label{sec:LCDM_dynamics}
\subsection{Deriving the Equations of Motion}
\label{sec:Derivation}

On the large scales for which a perturbative treatment is valid, we may approximate the cosmic matter density field as a single, self-gravitating, cold, pressureless fluid. Baryons follow the dark matter on these scales and pressure and other baryonic effects are negligible compared to gravitational interactions.
The matter density may therefore be characterised by its density contrast $\df(\vx,a)$ and peculiar velocity $\vv(\vx,a)$, where $\vx$ is comoving position and $a$ is the scale factor (we use $a$ and conformal time $\tau$, interchangeably, as time variables in the following). It is convenient to decompose the peculiar velocity into its divergence $\tf = \boldsymbol{\nabla}\cdot \vv (\vx,a)$ and vorticity $\vw = \boldsymbol{\nabla} \times \vv (\vx,a)$. These quantities obey the following equations of motion \citep{Fasiello+:2016,Desjacques:2018bnm}:
\eq{
\frac{\p\df_{\vk}}{\p\tau} + \tf_{\vk} &= -\int_{\vq_1,\vq_2}(2\pi)^3 \df^D_{\vk-\vq_{12}}\alpha(\vq_1,\vq_2)\tf_{\vq_1}\df_{\vq_2},\label{continuity} \\
\frac{\p\tf_{\vk}}{\p\tau} + \mathcal{H}\tf_{\vk} + \frac{3}{2}\Omega_m\mathcal{H}^2\df_{\vk} &= -\int_{\vq_1,\vq_2} (2\pi)^3 \df^D_{\vk-\vq_{12}}\beta(\vq_1,\vq_2)\tf_{\vq_1}\tf_{\vq_2},\label{euler}\\
\frac{\p\vw_{\vk}}{\p\tau} + \mathcal{H}\vw_{\vk}  &= i \vk \times \int _{\vq_1,\vq_2} (2\pi)^3 \df^D_{\vk-\vq_{12}} \vv_{\vq_1} \vw_{\vq_2} , \label{vorticity}
}
where $\delta_{\vq}^D$ is the Dirac delta function, $\vk_{12} = \vk_1 + \vk_2 $, $\int_{\vq_1,\ldots,\vq_n} \equiv (2\pi)^{-3n}\int d^3 \vq_1 \cdots d^3 \vq_n$, $\Omega_m(a)$ is the matter-density parameter and $\mathcal{H} = d\ln a/d \ln \tau$ is the conformal Hubble parameter.
The kernels are defined by $\alpha(\vq_1,\vq_2) = 1 + (\vq_1 \cdot \vq_2)/\vq_1^2$ and $\beta(\vq_1,\vq_2) = (\vq_{12})^2(\vq_1\cdot\vq_2)/(2\vq_1^2\vq_2^2)$. Equation~\eqref{vorticity} implies that $\vw$ remains zero for all time in the case of vanishing primordial vorticity, while it decays as $1/a$ at linear order if vorticity were present in the early universe. As a result, any small initial vorticity rapidly decays as the universe expands. Therefore, we proceed to assume the velocity field to be irrotational -- an assumption that holds well up to shell crossing (and the formation of shocks)~\citep{Bernardeau:2001qr}. 

The remaining system of Eqs.~(\ref{continuity}) and (\ref{euler}) are now closed and well-suited to be solved perturbatively, with an ansatz of
\eq{
	\df(\vk,a) &= \sum_{n = 1}^{\infty}\int_{\vq_1,\ldots,\vq_n} (2\pi)^3 \df^D_{\vk - \vq_{12...n}} F_n(\vq_1,\dots,\vq_n; a)D_+^n(a)\df^{\text{in}}_{\vq_1}\dots\df^{\text{in}}_{\vq_n},\nonumber\\
	\tf(\vk,a) &= -\mathcal{H}f_+(a)\sum_{n = 1}^{\infty}\int_{\vq_1,\ldots,\vq_n} (2\pi)^3 \df^D_{\vk - \vq_{12...n}} G_n(\vq_1,\dots,\vq_n; a)D_+^n(a)\df^{\text{in}}_{\vq_1}\dots\df^{\text{in}}_{\vq_n}.\label{ansatz}
}
Here, $F_n$ and $G_n$ are our solution kernels (defined to be fully symmetrized with respect to momenta), $D_{\pm}$ are the growing and decaying linear growth factors satisfying
\eq{
	a^2\frac{d^2 D_{\pm}}{da^2} + a\left(2 + \frac{d\ln\mathcal{H}}{d \ln a}\right)\frac{dD_{\pm}}{da} - \frac{3}{2}\Omega_m(a) D_{\pm} = 0,
}
in $\Lambda$CDM cosmologies, $f_{\pm} = d\ln D_{\pm} / d\ln a$ are the associated linear growth rates and $\df^{\mathrm{in}}_{\vq}$ describe the initial density contrast. At linear order, we find trivial results of $F_1 = G_1 = 1$, giving
\eeq{
\df_{\vk}^{(1)}(a) = D_+(a)\df^{\mathrm{in}}_{\vk}, \qquad \tf_{\vk}^{(1)} = - \mathcal{H}(a)f_+(a)D_+(a)\df^{\mathrm{in}}_{\vk} .
}

\begin{figure}[t!]
	\centering

    \resizebox{0.3\textwidth}{!}{\includegraphics{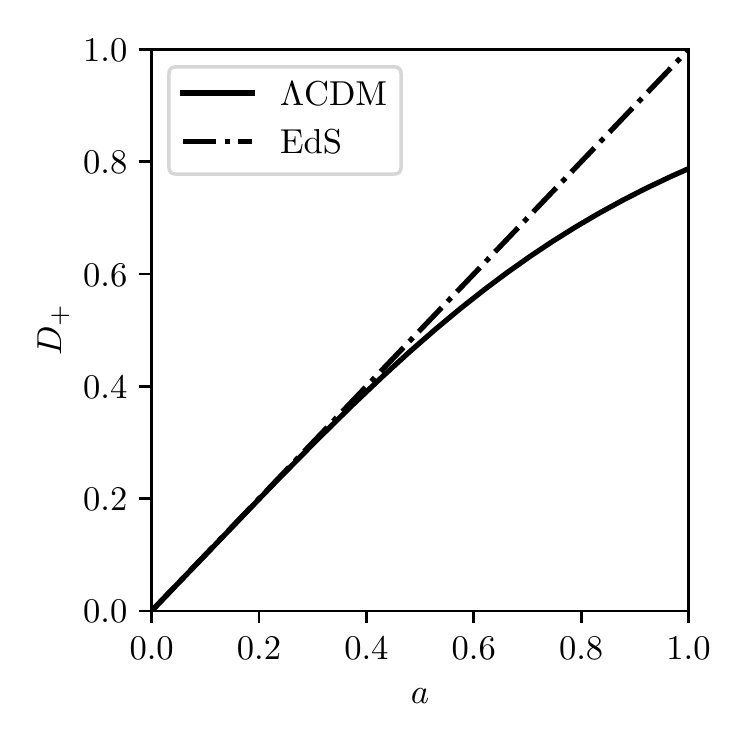}}
    \resizebox{0.3\textwidth}{!}{\includegraphics{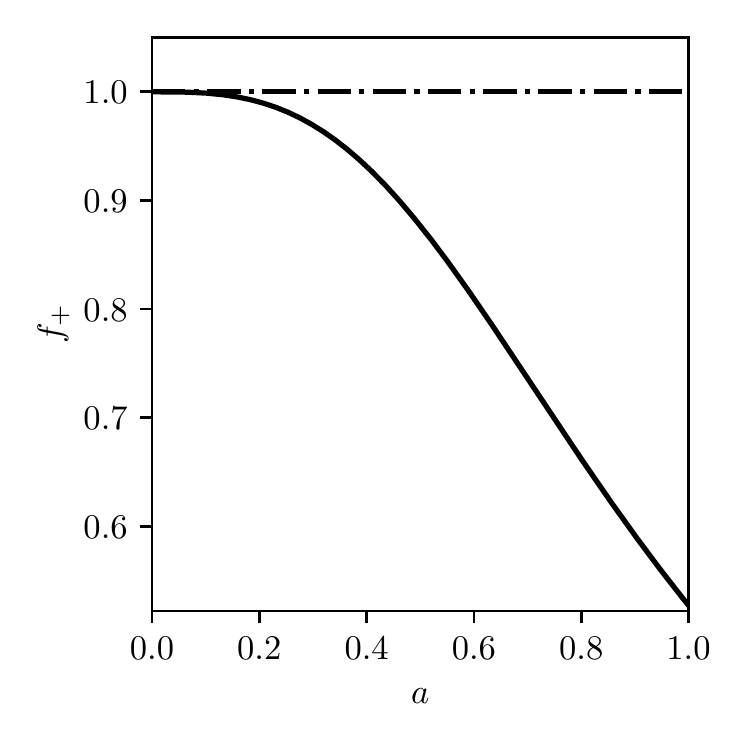}}
    \resizebox{0.3\textwidth}{!}{\includegraphics{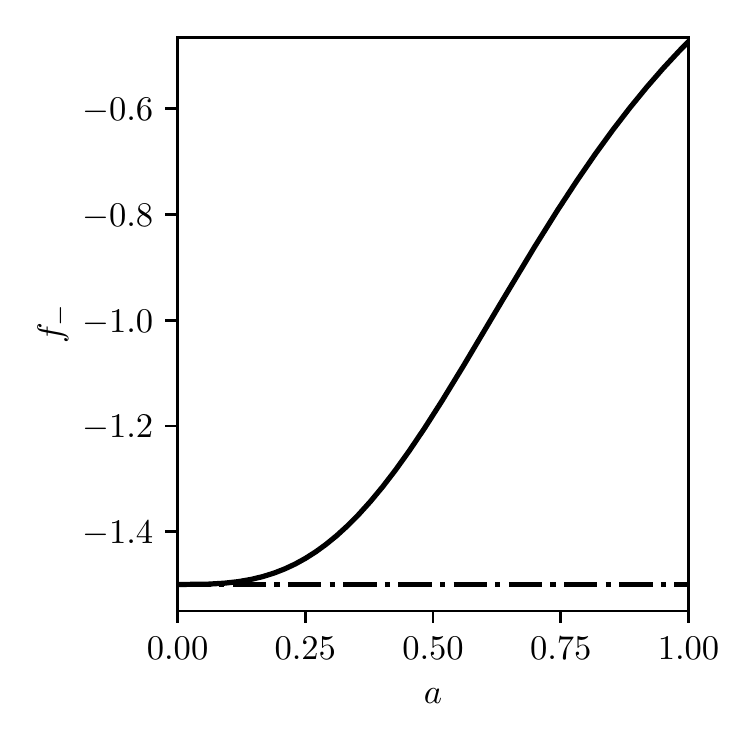}}
    
	\caption{\protect{\label{fig:growth}}Comparison of the growth factors for the growing mode (left), along with the logarithmic growth rates of the growing (middle) and decaying (right) modes, between the $\Lambda$CDM and quasi-EdS universes.}
\end{figure}

The linear growth factor for the growing mode, and its associated growth rate, are given by the following:
\begin{align}
D_+(a) &= \frac{5}{2}H_0^2\Omega_{m_0}H(a)\int_0^a\frac{dx}{[xH(x)]^3}, \label{eq:Dplus} \\
f_+(a) &= \frac{\Omega_{m_0}}{(1-\Omega_{m_0})a^3+\Omega_{m_0}}\left(\frac{5a}{2D_+(a)}-\frac{3}{2}\right) ,
\end{align}
where $H(a)=H_0[\Omega_{m_0}a^{-3} + (1-\Omega_{m_0})]^{1/2}$.
We have normalised $D_+(a)$ so that it approaches $a$ as $a \rightarrow 0$. The linear growth factor for the decaying mode is $D_-(a)\propto H(a)$ and the associated growth rate is
\eq{
f_-(a) = - \frac{3}{2}\frac{\Omega_{m_0}}{(1-\Omega_{m_0})a^3+\Omega_{m_0}} = -\frac{3}{2} \Omega_m(a).
}
Although we only consider growing-mode solutions, $f_-(a)$ appears below in the equation of motion for the $F_n$ and $G_n$ kernels through $d \ln H/d\ln a$, or, equivalently, $\Omega_m(a)$. Figure~\ref{fig:growth} shows the growth factor $D_{+}(a)$ along with the logarithmic growth rates $f_{\pm}$ in both the $\Lambda$CDM and EdS universes. In the latter, $D_+(a) = a$, $f_+(a)=1$ and $f_-(a)=-3/2$. The behaviour of these functions in $\Lambda$CDM differs from the EdS limits at late times and these deviations eventually compound to affect observables. The constancy of $f_{\pm}$ in EdS helps to explain how this approximation is viewed as ‘static’, as these growth rates do not introduce any dynamics into the system.
Typically, the quasi-EdS approximation is made, whereby the kernels $F_n$ and $G_n$ are assumed to be time-independent and are replaced by their counterparts in EdS, while the full $\Lambda$CDM growth factor ($D_{+}$) is used in Eq.~(\ref{ansatz}). This is done due to the fact that such kernels are much easier to compute in the simpler EdS case \citep{Bernardeau:2001qr, Fasiello+:2016, Fasiello:2016, Fasiello:2022lff}.

Dropping the qEdS approximation, these kernels obey the following equations of motion:
\begin{align}
    \frac{a}{f_+}\frac{d F_n}{d a} + nF_n - G_n = h_{\alpha}^{(n)}(\vq_1,...,\vq_n, a),\nonumber\\
    \frac{a}{f_+}\frac{d G_n}{d a} + (n-1)G_n - \frac{f_-}{f_+^2}(G_n - F_n) = h_{\beta}^{(n)}(\vq_1,...,\vq_n, a),
\end{align}
with source terms given by Eq.~(8) in \cite{Fasiello:2016}. In order to proceed, we assume a separable solution at each order, with ansatz \citep{Fasiello:2016, Fasiello:2022lff}
\begin{align}
F_n(\vq_1,...,\vq_n;a) &= \sum_{l = 1}^{N(n)}\lambda_n^{(l)}(a)\; H_n^{(l)}(\vq_1,...,\vq_n),\non\\
	G_n(\vq_1,...,\vq_n;a) &= \sum_{l = 1}^{N(n)}\kappa_n^{(l)}(a)\; H_n^{(l)}(\vq_1,...,\vq_n),
\end{align}
where $H_n^{(l)}$ are the momentum operators (see \cite{Fasiello:2022lff}) and $\lambda_n^{(l)}$ and $\kappa_n^{(l)}$ are the time-dependent coefficients, which are the focus of this paper. The numbering function $N(n)$ gives us the simplest way to ensure the full dimension of each kernel is accounted for, albeit allowing for some redundancies. These are imposed primarily through physical constraints such as conservation of mass and momentum, along with the equivalence principle --- though these are not explored here. The $N(n)$ are found recursively, with explicit form given in Appendix \ref{sec:appendixI}; the first few terms of which are $N(1)=1$, $N(2)=2$, $N(3)=6$, $N(4)=25$ and $N(5)=111$. For the purpose of this paper, we focus on the time-dependent coefficients, which can be written as~\citep{Fasiello:2022lff} 
\eq{
	\lambda_n^{(l)}(a) &= \df^K_{\frac{n}{2},\lfloor\frac{n}{2}\rfloor}\sum_{i = 1}^{N(n/2)}\bigg{[} \sum_{j = 1}^{N(n/2)}[W_{\alpha;\frac{n}{2},\frac{n}{2}}^{n(i,j)}]\df^K_{l,\phi_1} + \sum_{j = i}^{N(n/2)}[W_{\beta;\frac{n}{2},\frac{n}{2}}^{n(i,j)}]\df^K_{l,\phi_2}\bigg{]}\qquad\qquad\qquad\qquad\qquad\nonumber\\
	&\qquad+ \sum_{m = 1}^{\lfloor(n-1)/2\rfloor}\sum_{i = 1}^{N(m)}\sum_{j = 1}^{N(n - m)}\bigg{[}[W_{\alpha;m,n-m}^{n(i,j)}]\df^K_{l,\phi_3} + [W_{\alpha;n-m,m}^{n(j,i)}]\df^K_{l,\phi_4} + [W_{\beta;m,n-m}^{n(i,j)}]\df^K_{l,\phi_5} \bigg{]},\label{lambda} \\
	\kappa_n^{(l)}(a) &= \df^K_{\frac{n}{2},\lfloor\frac{n}{2}\rfloor}\sum_{i = 1}^{N(n/2)}\bigg{[} \sum_{j = 1}^{N(n/2)}[U_{\alpha;\frac{n}{2},\frac{n}{2}}^{n(i,j)}]\df^K_{l,\phi_1} + \sum_{j = i}^{N(n/2)}[U_{\beta;\frac{n}{2},\frac{n}{2}}^{n(i,j)}]\df^K_{l,\phi_2}\bigg{]}\qquad\qquad\qquad\qquad\qquad\nonumber\\
	&\qquad+ \sum_{m = 1}^{\lfloor(n-1)/2\rfloor}\sum_{i = 1}^{N(m)}\sum_{j = 1}^{N(n - m)}\bigg{[}[U_{\alpha;m,n-m}^{n(i,j)}]\df^K_{l,\phi_3} + [U_{\alpha;n-m,m}^{n(j,i)}]\df^K_{l,\phi_4} + [U_{\beta;m,n-m}^{n(i,j)}]\df^K_{l,\phi_5} \bigg{]},\label{kappa}
}
for $n>1$. For the case $n=1$, we have $\lambda^{(1)}_1 = \kappa^{(1)}_1 = 1$. In these expressions, $\df^K$ is the Kronecker delta and $\phi_k$ are bijective maps of their arguments ($n$ and all of the integers being summed over) that serve to identify the correct coefficient for each momentum operator. Their form is given explicitly in Appendix \ref{sec:appendixI}. For $\lambda_n^{(l)}$, exactly one of the $W_\alpha(a)$ or $W_\beta(a)$ coefficients are selected, while for $\kappa_n^{(l)}$ it is one of the $U_\alpha(a)$ or $U_\beta(a)$.
Though this formulation appears terse, we reiterate that this represents an algorithm that is capable of readily producing all information necessary to each perturbative order. For pedagogical reasons, we summarise the amount of information necessary to predict typical observables in Tab.~\ref{table:observables} \cite{Sefusatti:2011cm, Bose:2018zpk}.
\begin{table}[ht!]
	\begin{center}
		\begin{tabular}{c c c }
			\hline
            \hline
			\textbf{Required order ($n$)} & \textbf{Observable} & \textbf{Number of dynamical coefficients} \\
			\hline
			1 & Linear power spectrum & 2 (trivial) \\
			2 & Tree-level bispectrum & 6\\
			3 & One-loop power spectrum & 18\\
			4 & One-loop bispectrum & 68\\
            5 & Two-loop power spectrum & 290\\
			\hline
		\end{tabular}
	\caption{\protect{\label{table:observables}}Cumulative information required to calculate typically used observables of the matter and velocity fluctuations. The number of coefficients required is calculated as $2\sum_{i = 1}^{n}N(i)$.}
	\end{center}
\end{table}

The $W$ and $U$ functions satisfy coupled differential equations, with the following structure:
\eq{
	\dot{W}_{\alpha;m_1,m_2}^{n(i,j)} + nW_{\alpha;m_1,m_2}^{n(i,j)} - U_{\alpha;m_1,m_2}^{n(i,j)} &= \kappa_{m_1}^{(i)}\lambda_{m_2}^{(j)},\nonumber\\
	\dot{U}_{\alpha;m_1,m_2}^{n(i,j)} + (n-1)U_{\alpha;m_1,m_2}^{n(i,j)} - \frac{f_-}{f_+^2}\bigg{[}U_{\alpha;m_1,m_2}^{n(i,j)} - W_{\alpha;m_1,m_2}^{n(i,j)}\bigg{]} &= 0,\label{alpha}\\
	\dot{W}_{\beta;m_1,m_2}^{n(i,j)}+ nW_{\beta;m_1,m_2}^{n(i,j)} - U_{\beta;m_1,m_2}^{n(i,j)} &= 0, \nonumber\\
	\dot{U}_{\beta;m_1,m_2}^{n(i,j)} + (n-1)U_{\beta;m_1,m_2}^{n(i,j)} - \frac{f_-}{f_+^2}\bigg{[}U_{\beta;m_1,m_2}^{n(i,j)} - W_{\beta;m_1,m_2}^{n(i,j)} \bigg{]} &= \kappa_{m_1}^{(i)}\kappa_{m_2}^{(j)},\label{beta}
}
where $n = m_1 + m_2$, and overdots denote differentiation with respect to $\eta$, where $\eta = \ln D_+$ is used as a reduced time coordinate, so that
$d/d\eta  = (a/f_+)d/da$. Equations~\eqref{alpha} and~\eqref{beta} have initial conditions as $a\rightarrow 0$ given by their time-independent EdS values, found using the recursion relations
\eq{
    [W_{\alpha;m_1,m_2}^{n(i,j)}]^{\mathrm{EdS}} &=  \bigg{(}\frac{2n + 1}{2n^2 + n - 3} \bigg{)}[\kappa_{m_1}^{(i)}]^{\mathrm{EdS}}[\lambda_{m_2}^{(j)}]^{\mathrm{EdS}},\non\\
    [U_{\alpha;m_1,m_2}^{n(i,j)}]^{\mathrm{EdS}} &= \bigg{(}\frac{3}{2n^2 + n - 3} \bigg{)}[\kappa_{m_1}^{(i)}]^{\mathrm{EdS}}[\lambda_{m_2}^{(j)}]^{\mathrm{EdS}},\label{alphaEDS}\\
    [W_{\beta;m_1,m_2}^{n(i,j)}]^{\mathrm{EdS}} &= \bigg{(}\frac{2}{2n^2 + n - 3} \bigg{)}[\kappa_{m_1}^{(i)}]^{\mathrm{EdS}}[\kappa_{m_2}^{(j)}]^{\mathrm{EdS}} ,\non\\
    [U_{\beta;m_1,m_2}^{n(i,j)}]^{\mathrm{EdS}} &= \bigg{(}\frac{2n}{2n^2 + n - 3} \bigg{)}[\kappa_{m_1}^{(i)}]^{\mathrm{EdS}}[\kappa_{m_2}^{(j)}]^{\mathrm{EdS}} .\label{betaEDS}
}
The efficient solution of the equations of motion~(\ref{alpha}) and (\ref{beta}) is the main goal of this paper, as the iterative source terms make their numerical evaluation relatively computationally intensive. This is accentuated by the information in Tab. \ref{table:observables}, which shows just how many of these functions must be calculated to predict each observable. As a result, this work aims to find an efficient method to determine the solution of non-linear, coupled ordinary differential equations of this form. Doing so will allow us to take into account efficiently the full time-dependence of the non-linear density and velocity fields in perturbation-theory calculations.

\subsection{Direct Numerical Solution for the Equations of Motion}
\label{sec:numerical}

To establish a benchmark, we first solve~Eqs. (\ref{lambda}) and (\ref{kappa}) directly with a basic LSODA solver for first-order ordinary differential equations (ODEs)~\citep{Hindermarsh:1983,Petzold:1983}. We use initial conditions defined in Eqs. (\ref{alphaEDS}) and (\ref{betaEDS}), giving for instance 
\eq{
[W_{\alpha;1,1}^{2(1,1)}]^{\mathrm{EdS}} = \frac{5}{7}, \;\;\;[U_{\alpha;1,1}^{2(1,1)}]^{\mathrm{EdS}} = \frac{3}{7},\qquad\qquad\;\;\;\;\non\\
[W_{\beta;1,1}^{2(1,1)}]^{\mathrm{EdS}} = \frac{2}{7},\;\;\; [U_{\beta;1,1}^{2(1,1)}]^{\mathrm{EdS}} = \frac{4}{7},\;\;\;\mathrm{for}\;n = 2. \non
}
This process was completed up to third-order, with second-order solutions being used in turn to generate higher-order results. Figure~\ref{fig:numerical} shows the solutions for $\lambda_3^{(l)}$ and $\kappa_3^{(l)}$ for $l=1$--$6$ (since $N(3)=6$),
found using this method. These have been normalised by their respective EdS values in order to make visual comparison easier. We find that in the majority of cases these coefficients tend to depart from EdS with increasing $a$. It is important to note that the iterative nature of Eqs.~(\ref{alpha}) and (\ref{beta}) makes this method progressively more time-consuming as we aim to solve for higher $n$, corresponding to higher-order dynamics. Therefore, a more efficient method of solution is clearly desirable.

\begin{figure}[ht!]
	\centering
	\begin{subfigure}[b]{0.475\textwidth}
		\centering
        \includegraphics{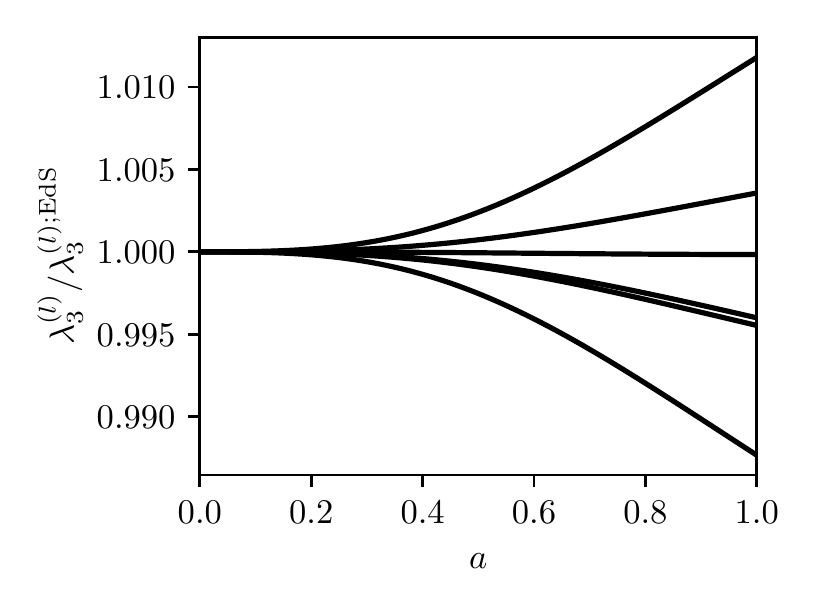}
		\caption{$\lambda_3^{(l)}$}   
	\end{subfigure}
	\hfill
	\begin{subfigure}[b]{0.475\textwidth}
		\centering 
        \includegraphics{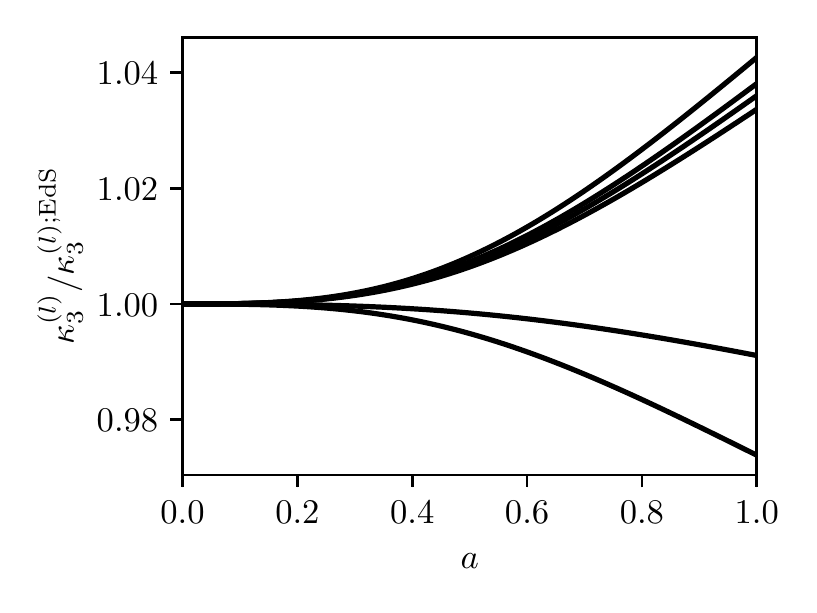}
		\caption{$\kappa_3^{(l)}$}
	\end{subfigure}
	\caption{\protect{\label{fig:numerical}}Numerical solutions for $\lambda_n^{(l)}$ (left) and $\kappa_n^{(l)}$ (right) for $n = 3$ obtained from Eqs.~\eqref{lambda} and~\eqref{kappa} with the ODEs~\eqref{alpha} and~\eqref{beta} solved by direct numerical integration. The solutions are normalised by their respective EdS values. These coefficients start to depart from their EdS values with increasing $a$, reaching around 1\% and 4\% differences for $\lambda_n^{(l)}$ and $\kappa_n^{(l)}$, respectively.}
\end{figure}

\section{The Chebyshev Spectral Method}
\label{sec:spectral}

As stated above, we aim to find an efficient method to solving Eqs. (\ref{lambda}--\ref{beta}) in a full $\Lambda$CDM universe. We do so using the Chebyshev spectral method (CSM), based around an expansion in Chebyshev polynomials. A code implementation of the CSM in Python is available at \url{https://github.com/Chousti/CSMethod.git}, with its structure presented in Sec.~\ref{sec:csm}.

\subsection{Shifted Chebyshev Polynomials}
\label{sec:polynomials}

Before we proceed to describe the CSM, we introduce the Chebyshev polynomials themselves \citep{Chebyshev:1853}. These are two families of polynomials defined in relation to trigonometric functions: 
\begin{align}
	T_n(\cos\tf) &= \cos (n\tf), \nonumber \\
	U_n(\cos\tf)\sin\tf &= \sin(n+1)\tf,
\end{align}
typically defined in the domain $\tf \in [-\pi/2,\pi/2]$. These are referred to as the Chebyshev polynomials of the first and second kind, respectively. We use the first kind, which can also be generated from the recurrence relation
\eeq{
	T_0 = 1, \qquad T_1 = x, \qquad T_{n+1}(x) = 2xT_n(x) - T_{n-1}(x),
 }
where $x \in [-1,1]$.

These polynomials are particularly useful as they are orthogonal within their domain with respect to the weight function $w(x) = (1 - x^2)^{-1/2}$. This is because they are solutions to the Chebyshev differential equations, which are of Sturm--Liouville form.

In order to make the Chebyshev polynomials suitable for our problem, we must rescale them, such that we can use the cosmological scale factor $a \in [0,1]$ as the argument. This is done by defining the shifted Chebyshev polynomials,
\eeq{
\tilde{T}_n(x) = T_n\left(\frac{2x - c - b}{c - b}\right),
}
where, for generality, we have considered the domain $[b,c]$. We now proceed to outline several key properties (for this arbitrary shift) which will be particularly useful. These include their orthogonality relations:
\eeq{
	\int_b^c \tilde{T}_n(x)\tilde{T}_m(x) \frac{dx}{\sqrt{x(b + c - x) - bc}} = \begin{cases}
		\mbox{0} & \mbox{if }  n \neq m,\\
		\pi & \mbox{if } n = m = 0,\\
		\frac{\pi}{2} & \mbox{if } n = m \neq 0;
	\end{cases}
}
product relations:
\eeq{\label{product}
	2\tilde{T}_n(x)\tilde{T}_m(x) = \tilde{T}_{n+m}(x) + \tilde{T}_{|n - m|}(x);
}
and integral relation:
\begin{equation}
 \frac{4}{c-b} \int \tilde{T}_n(x) \, dx = \frac{\tilde{T}_{n+1}(x)}{n+1} - \frac{\tilde{T}_{n-1}(x)}{n-1} \qquad (n\geq 2)\, .  
 \label{eq:intrelation}
\end{equation}
These properties were derived based on known results for the original polynomials. For the full derivations of the original properties, along with a myriad of others, the reader may consult \citep{Karjanto:2020}.

It is now possible to approximate an arbitrary, smooth function $y(x)$, valid in the range $[b,c]$, with a truncated sum of shifted polynomials:
\eeq{\label{expansion}
	y(x) = \sum_{i = 0}^{N}a_i\tilde{T}_i(x) = \vec{a}\cdot\vec{t},
}
where the expansion coefficients are given by
\eeq{\label{expansionterm}
	a_i = \frac{c_i}{\pi}\int_b^c y(u)\tilde{T}_i(u) \frac{du}{\sqrt{u(b + c - u) - bc}}, \; \; \; c_i = \begin{cases}
		2 & \mbox{if }  i \neq 0,\\
		1 & \mbox{if }  i = 0.
	\end{cases}
}
Henceforth, we will refer to $\vec{a}$ as the vector of components of $y(x)$, of length $N$. In this vectorial language, it is possible to express operations such as derivatives and products of functions as matrix operations. Namely, we have the derivative
\eeq{
	\frac{dy}{dx} = \vec{a}^{\prime}\cdot\vec{t} = (\vec{D}\cdot\vec{a})\cdot\vec{t} \; \rightarrow \; \vec{a}^{\prime} = \vec{D}\cdot\vec{a},
}
where we have defined
\eeq{
	\vec{D} = \frac{4}{c - b} \begin{bmatrix} 
		0 & \frac{1}{2} & 0 & \frac{3}{2} & 0 & \frac{5}{2} &\cdots\\
		0 & 0 & 2 & 0 & 4 & 0 \\
		0 & 0 & 0 & 3 & 0 & 5 \\
		0 & 0 & 0 & 0 & 4 & 0 \\
		0 & 0 & 0 & 0 & 0 & 5 \\
		\vdots & & & & & & \ddots 
	\end{bmatrix}.
}
This follows from integrating $dy/dx = \vec{a}^{\prime}\cdot\vec{t}$ and using Eq.~\eqref{eq:intrelation}. Similarly, it is possible to expand the product of two functions using Eq. (\ref{product}) as
\eeq{
	y(x) = g(x)h(x) = (\vec{g}\cdot\vec{t})(\vec{h}\cdot\vec{t}) = (\vec{P}(\vec{g})\cdot\vec{h})\cdot\vec{t} \; \rightarrow \; \vec{a} = \vec{P}(\vec{g})\cdot\vec{h}, \label{Product}
}
where we have defined
\eeq{
	\vec{P}(\vec{g}) = \frac{1}{2} \begin{bmatrix} 
		2g_0 & g_1 & g_2 & g_3 & \cdots\\
		2g_1 & 2g_0 + g_2 & g_1 + g_3 & g_2 + g_4  &\\
		2g_2 & g_1 + g_3 & 2g_0 + g_4 & g_1 + g_5 & \\
		2g_3 & g_2 + g_4 & g_1 + g_5 & 2g_0 + g_6  & \\
		\vdots & & & & \ddots 
	\end{bmatrix}.
}
Here, we note that two different $\vec{P}$ matrices commute between each other (i.e., $\vec{P}(\vec{g})\cdot \vec{h} = \vec{P}(\vec{h})\cdot \vec{g}$). Furthermore, $\vec{P}$ and $\vec{D}$ matrices will not commute, providing a useful sanity check. This formalism will prove particularly useful and will be implemented fully in Section \ref{sec:csm}. Though these results are given for an arbitrary rescaling of the argument, henceforth we shall proceed with $\tilde{T}$ denoting a shift to the domain $[0,1]$.

Given the ODEs~\eqref{alpha} and~\eqref{beta} that we need to solve, when expressed in terms of derivatives with respect to $a$ we shall require decompositions of the functions $f_-/f_+^2$ and $1/f_+$ into the polynomial basis. To allow for future generalisability, this is done numerically using Eqs.~(\ref{expansion}) and (\ref{expansionterm}) with the accuracy of the recomposition for $f_-/f_+^2$ shown in Fig.~\ref{fig:fdecomp} for $N \in [2,4,6]$ Chebyshev components. Here we find that using $N = 4$ yields sub-percent-level accuracy across the whole function. The same is true for the decomposition of $1/f_+$, but this is not shown for brevity.

\begin{figure}[ht!]
	\centering
	\begin{subfigure}[b]{0.9\textwidth}
		\centering
        \includegraphics{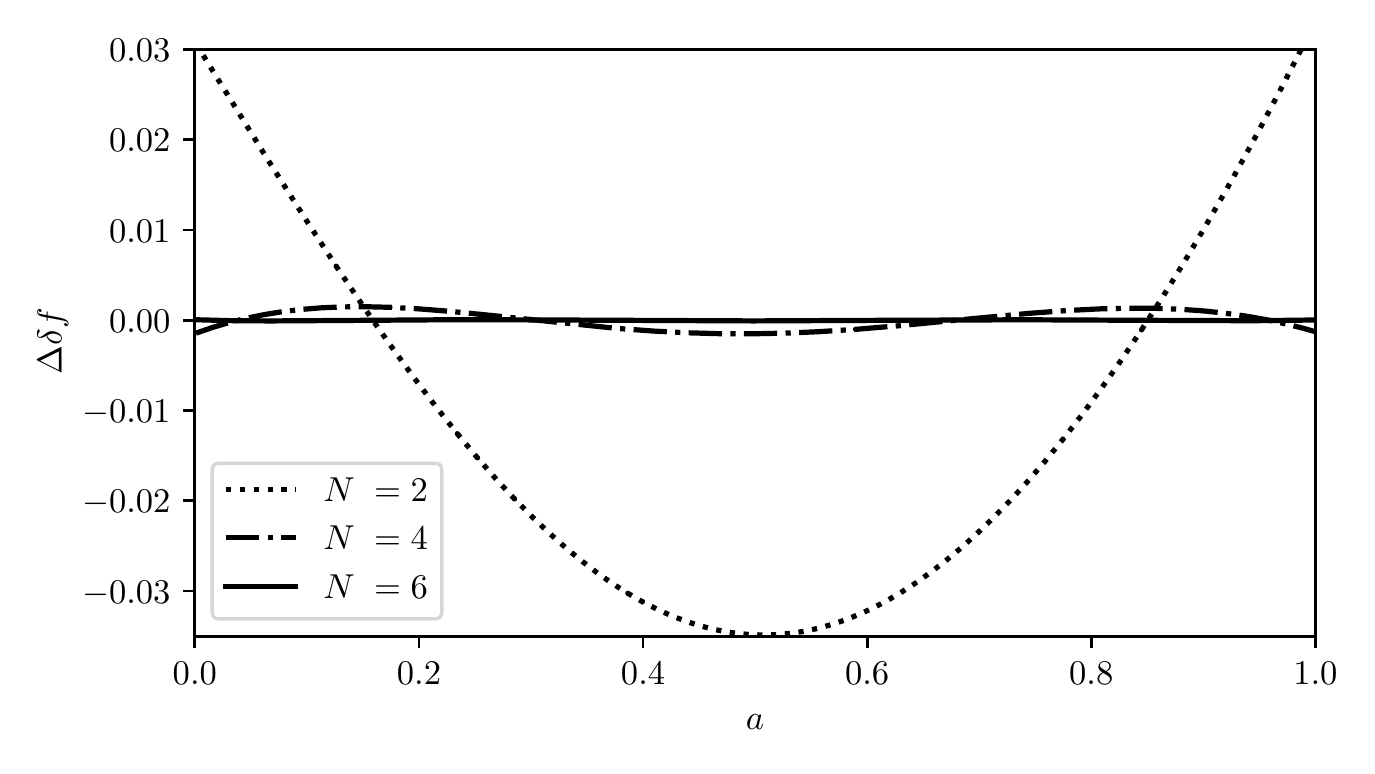}
	\end{subfigure}
	
	\vskip\baselineskip
	\begin{subfigure}[b]{0.9\textwidth}
		\centering 
        \includegraphics{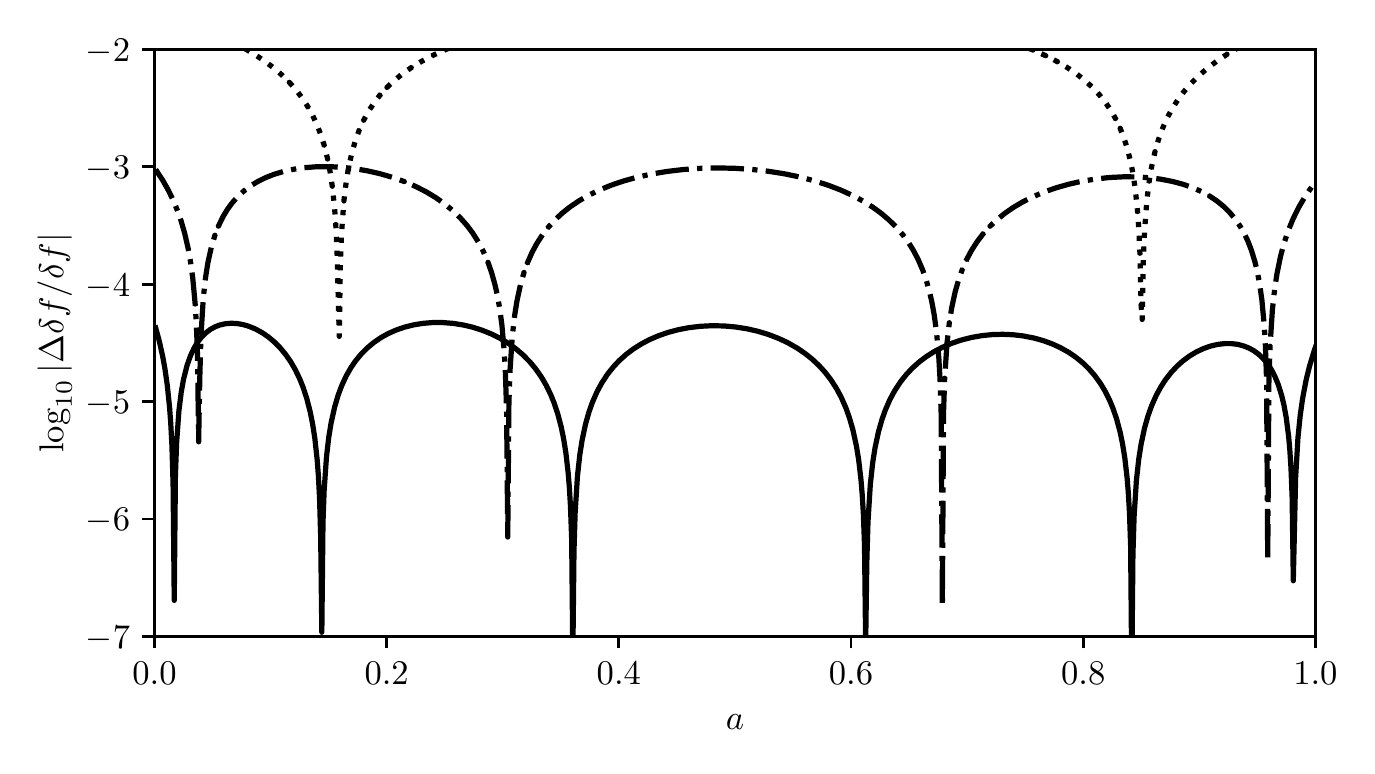}
	\end{subfigure}
	\caption{Numerical decomposition of $\df f = f_-/f_+^2 + 3/2$ into the shifted Chebyshev polynomial basis. Both the residue (top) and absolute values of the relative error (bottom) are shown. The residue is given by $\Delta X = X - X_{\mathrm{true}}$.}
    \label{fig:fdecomp}
\end{figure}

\subsection{Implementation of CSM}
\label{sec:csm}

We now discuss the implementation and structure of the CSM. Generally, spectral methods work by numerically determining the coefficients in a given basis of a differential equation subject to boundary conditions \citep{Boyd:2000}. Particularly, the CSM makes use of Chebyshev polynomials because they are easy to compute and rapidly convergent as compared to other basis functions such as the Legendre Polynomials \citep{Boyd:2000,Clenshaw:1957,Sezer:1996}.

On the surface, this method works by using the polynomials' properties to convert differential equations into matrix equations, corresponding to a system of linear equations for a set of unknown components, with the number of components ($N$) corresponding to the user's desired accuracy. Therefore, the problem's complexity has been reduced to one of simple linear algebra, for which a variety of optimised codes exist. As a result, the method will naively always be faster than one using numerical integration routines. 

We proceed by defining the following expansions, using the formalism outlined in Sec.~\ref{sec:polynomials} of shifted Chebyshev polynomials:
\eq{
    \lambda(a) &= \vec{L}\cdot\vec{t}(a), \; \; \kappa(a) = \vec{K}\cdot\vec{t}(a), \label{expansionsol}\\
    W(a) &= \vx\cdot\vec{t}(a), \; \; U(a) = \vec{y}\cdot\vec{t}(a),\\
	\frac{f_-}{f_+^2}(\Omega_{m_0},a) &= \vec{d}(\Omega_{m_0})\cdot\vec{t}(a), \; \; f_+^{-1}(\Omega_{m_0},a) = \vec{c}(\Omega_{m_0})\cdot\vec{t}(a),\;\; a = \vec{e}\cdot\vec{t}(a),
}
where $\vec{e} = (1,0,0,\ldots)$ and all subscripts and superscripts have been implied but omitted for brevity. The components $c$ and $d$ are calculated as shown in Fig. \ref{fig:fdecomp} on the first iteration of each run.

Next, we substitute these relations into Eqs.~(\ref{lambda}) and (\ref{kappa}) for a given cosmology and utilise properties of the polynomials to find:
\eq{
	\vec{L}_n^{(l)} = \df^K_{\frac{n}{2},\lfloor\frac{n}{2}\rfloor}\sum_{i = 1}^{N(n/2)}\bigg{[} \sum_{j = 1}^{N(n/2)}[\vx_{\alpha}]^{n(i,j)}_{\frac{n}{2},\frac{n}{2}}\df^K_{l,\phi_1} + \sum_{j = i}^{N(n/2)}[\vx_{\beta}]^{n(i,j)}_{\frac{n}{2},\frac{n}{2}}\df^K_{l,\phi_2}\bigg{]}\qquad\qquad\qquad\qquad\qquad\non\\
	+ \sum_{m = 1}^{\lfloor(n-1)/2\rfloor}\sum_{i = 1}^{N(m)}\sum_{j = 1}^{N(n - m)}\bigg{[}[\vx_{\alpha}]^{n(i,j)}_{m,n-m}\df^K_{l,\phi_3} + [\vx_{\alpha}]^{(ji)}_{n-m,m}\df^K_{l,\phi_4} + [\vx_{\beta}]^{n(i,j)}_{m,n-m}\df^K_{l,\phi_5} \bigg{]}\label{subs1},
}
\eq{
	\vec{K}_n^{(l)} = \df^K_{\frac{n}{2},\lfloor\frac{n}{2}\rfloor}\sum_{i = 1}^{N(n/2)}\bigg{[} \sum_{j = 1}^{N(n/2)}[\vec{y}_{\alpha}]^{n(i,j)}_{\frac{n}{2},\frac{n}{2}}\df^K_{l,\phi_1} + \sum_{j = i}^{N(n/2)}[\vec{y}_{\beta}]^{n(i,j)}_{\frac{n}{2},\frac{n}{2}}\delta^K_{l,\phi_2}\bigg{]}\qquad\qquad\qquad\qquad\qquad\non\\
	+ \sum_{m = 1}^{\lfloor(n-1)/2\rfloor}\sum_{i = 1}^{N(m)}\sum_{j = 1}^{N(n - m)}\bigg{[}[\vec{y}_{\alpha}]^{n(i,j)}_{m,n-m}\df^K_{l,\phi_3} + [\vec{y}_{\alpha}]^{(ji)}_{n-m,m}\df^K_{l,\phi_4} + [\vec{y}_{\beta}]^{n(i,j)}_{m,n-m}\df^K_{l,\phi_5} \bigg{]}\label{subs2},
}
where the coupled system of ODEs~(\ref{alpha}) and (\ref{beta}) becomes a system of algebraic equations:
\eq{
	\bigg{[}\vec{P}(\vec{c}) \cdot \vec{P}(\vec{e}) \cdot \vec{D} + n\vec{I}\bigg{]}\cdot [\vec{x}_{\alpha}]_{m_1,m_2}^{n(i,j)} - \vec{I}\cdot [\vec{y}_{\alpha}]_{m_1,m_2}^{n(i,j)} &= \vec{P}(\vec{K}_{m_1}^{(i)})\cdot\vec{L}_{m_2}^{(j)},\non\\
	\vec{P}(\vec{d}) \cdot [\vec{x}_{\alpha}]_{m_1,m_2}^{n(i,j)} + \bigg{[} \vec{P}(\vec{c}) \cdot \vec{P}(\vec{e}) \cdot \vec{D} + (n - 1)\vec{I} - \vec{P}(\vec{d})\bigg{]}\cdot[\vec{y}_{\alpha}]_{m_1,m_2}^{n(i,j)} &= 0 \label{expalpha},
}
and
\eq{
	\bigg{[}\vec{P}(\vec{c}) \cdot \vec{P}(\vec{e}) \cdot \vec{D} + n\vec{I}\bigg{]}\cdot [\vec{x}_{\beta}]_{m_1,m_2}^{n(i,j)} - \vec{I}\cdot [\vec{y}_{\beta}]_{m_1,m_2}^{n(i,j)} &= 0,\non\\
	\vec{P}(\vec{d}) \cdot [\vec{x}_{\beta}]_{m_1,m_2}^{n(i,j)} + \bigg{[} \vec{P}(\vec{c}) \cdot \vec{P}(\vec{e}) \cdot \vec{D} + (n - 1)\vec{I} - \vec{P}(\vec{d})\bigg{]}\cdot[\vec{y}_{\beta}]_{m_1,m_2}^{n(i,j)} &= \vec{P}(\vec{K}_{m_1}^{(i)})\cdot\vec{K}_{m_2}^{(j)}\label{expbeta}
}
for the $\alpha$ and $\beta$ systems, respectively. It is important to note several aspects. First, the length ($N+1$) of the unknown vectors $\vx$ and $\vec{y}$ corresponds to the greatest order of Chebyshev polynomial used ($N$) and is set by the user, with the implications discussed in Section \ref{sec:spectraltest}. Secondly, due to the iterative nature of these equations, in principle, the source terms on the right-hand sides of Eqs. (\ref{expalpha}) and (\ref{expbeta}) are just vectors of constants, which we shall henceforth define as $\vec{\sigma}_{\alpha/\beta}$ and $\vec{\tau}_{\alpha/\beta}$, respectively. We note that $\vec{\tau}_{\alpha}=\vec{0}$ and $\vec{\sigma}_{\beta}=\vec{0}$.
These source terms are calculated recursively, making use of the code's inherent structure to ensure this is done efficiently. Finally, one small approximation is made to make the code even more efficient. In particular, it was noted that for all functions decomposed in this way, the components $a_i$ were $\mathcal{O}(10^{-i})$. Therefore, in Eq.~(\ref{expansion}), the code forces all components of orders greater than $N$ to zero.

Schematically, we can simplify Eqs. (\ref{expalpha}) and (\ref{expbeta}) into matrix equations of dimension $2N+2$:
\eq{
	\left(\begin{array}{@{}c|c@{}}
		\vec{E} & \vec{F} \\\hline
		\vec{G} & \vec{H} \\
	\end{array}\right) 
	\left(\begin{array}{@{}c@{}}
		\vec{x}_{\mu;m_1,m_2}^{n(i,j)} \\
		\vec{y}_{\mu;m_1,m_2}^{n(i,j)}
	\end{array}\right) = 
	\left(\begin{array}{@{}c@{}}
		\vec{\sigma}_{\mu;m_1,m_2}^{n(i,j)} \\
		\vec{\tau}_{\mu;m_1,m_2}^{n(i,j)}
	\end{array}\right)\label{linalg},
}
where $\mu = \alpha, \beta$ denotes which system is being computed.
We must also implement our boundary conditions, namely that our dynamical coefficients reduce to their EdS counterparts as $a \rightarrow 0$. Component-wise, this becomes for $W$:
\eq{
[W_{\mu;m_1,m_2}^{n(i,j)}]^{\mathrm{EdS}} = \sum_{k=0}^{N}[x_{\mu;m_1,m_2}^{n(i,j)}]_k\tilde{T}_k(0) = \sum_{k=0}^{N}(-1)^k[x_{\mu;m_1,m_2}^{n(i,j)}]_k,
}
with a similar result for $U$. We then force this constraint on the system by replacing the bottom row of $\vec{E}$ with $[1,-1,...,(-1)^{N+1}]$, the bottom row of $\vec{F}$ with an ($N+1$)-tuple of zeros, and finally setting $[\sigma_{\mu;m_1,m_2}^{n(i,j)}]_N = [W_{\mu;m_1,m_2}^{n(i,j)}]^{\mathrm{EdS}}$. We then repeat this process for $\vec{H}$, $\vec{G}$ and  $\vec{\tau}_{\mu;m_1,m_2}^{n(i,j)}$, respectively.

Therefore, we have successfully reduced our system of coupled ODEs to a basic matrix multiplication problem, for which there exist a variety of efficient linear algebra methods. For the purposes of the CSM, a well-established LAPACK routine was used \citep{Anderson:1999} to invert Eq. (\ref{linalg}). Once these unknown components have been found, we can recompose them to find $W$ and $U$. In turn, Eqs. (\ref{subs1}) and (\ref{subs2}) can then be used to find the dynamical solutions for Eqs. (\ref{lambda}) and (\ref{kappa}), therefore solving our problem.

\section{Results and comparison of solutions}
\label{sec:results}
\subsection{Testing the Chebyshev Spectral Method}
\label{sec:spectraltest}

The method described in Section \ref{sec:spectral} has been implemented, and is here tested by means of solving Eqs. (\ref{lambda}) and (\ref{kappa}) for the case of a $\Lambda$CDM universe with initial conditions given by the EdS limit (Eqs.~\ref{alphaEDS} and~\ref{betaEDS}) and with $\Omega_{m_0} = 0.315$. This case was chosen for illustrative purposes only, as the CSM is capable of efficiently producing results for any $\Lambda$CDM universe, allowing it to be used to scan across the full parameter space of $\Omega_{m_0}$.

Figure~\ref{fig:all} shows all solutions to Eq. (\ref{lambda}) for the $\lambda_n^{(l)}$ (red) and Eq.~(\ref{kappa}) for the $\kappa^{(l)}_n$  (blue) for $n\in[2,3,4,5]$ and for all associated values of $l$ as generated by the CSM with $N = 5$. This demonstrates both the increasing number of dynamical functions and their greater deviation from the EdS limit with increased perturbative order. These functions represent all those necessary in order to derive the matter-density and associated velocity power spectra to two-loop-order, as given in \citep{Fasiello:2022lff}.

\begin{figure}[ht!]
	\centering
	\begin{subfigure}[b]{0.475\textwidth}
		\centering
        \includegraphics{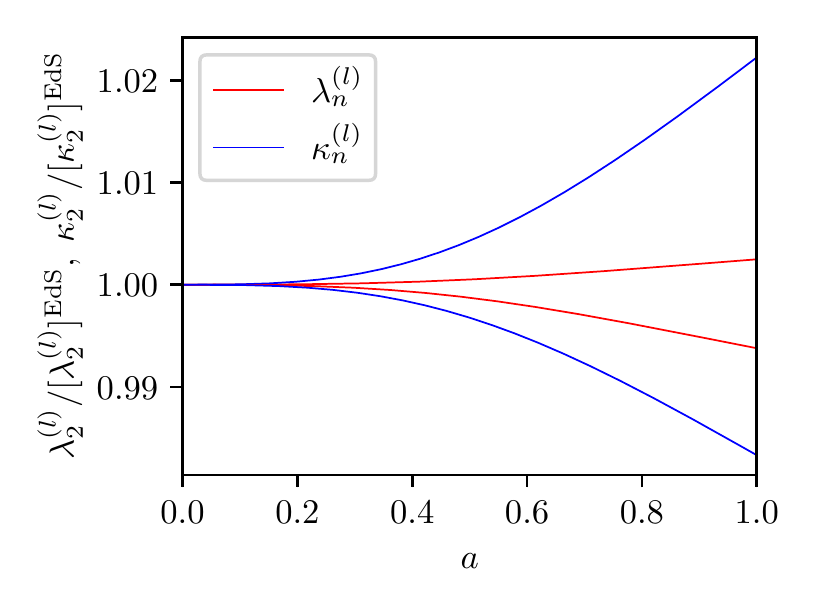}
		\caption{$n = 2$}   
	\end{subfigure}
	\hfill
	\begin{subfigure}[b]{0.475\textwidth}
		\centering 
        \includegraphics{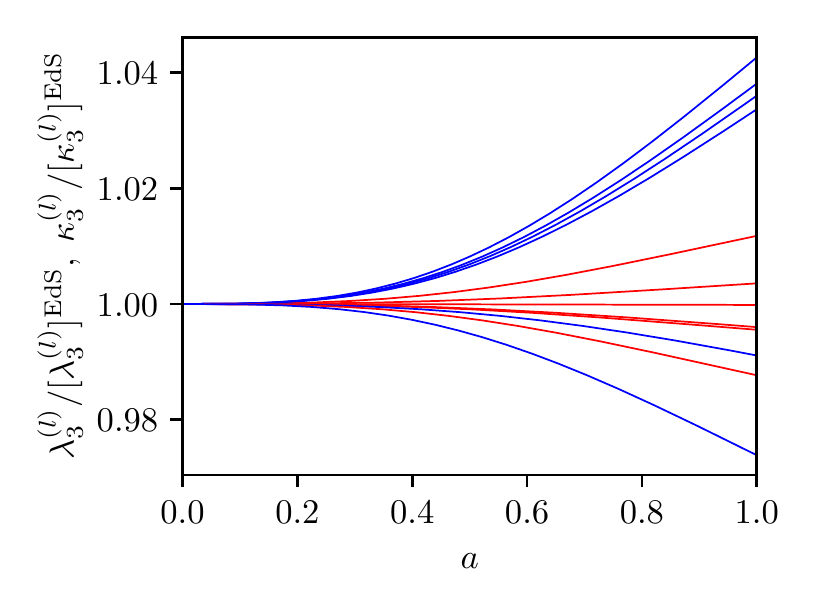}
		\caption{$n = 3$}
	\end{subfigure}
	\vskip\baselineskip
	\begin{subfigure}[b]{0.475\textwidth}
		\centering 
        \includegraphics{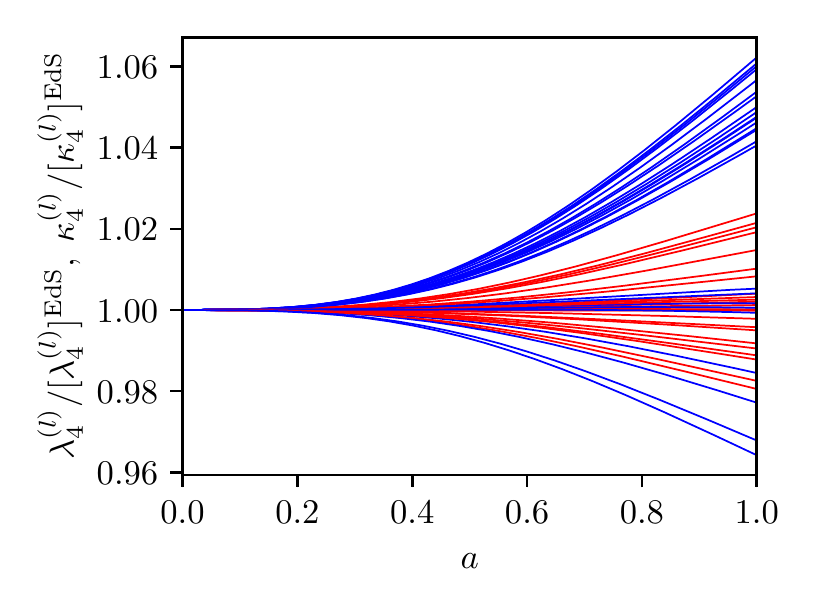}
		\caption{$n = 4$}
	\end{subfigure}
	\hfill
	\begin{subfigure}[b]{0.475\textwidth}   
		\centering 
        \includegraphics{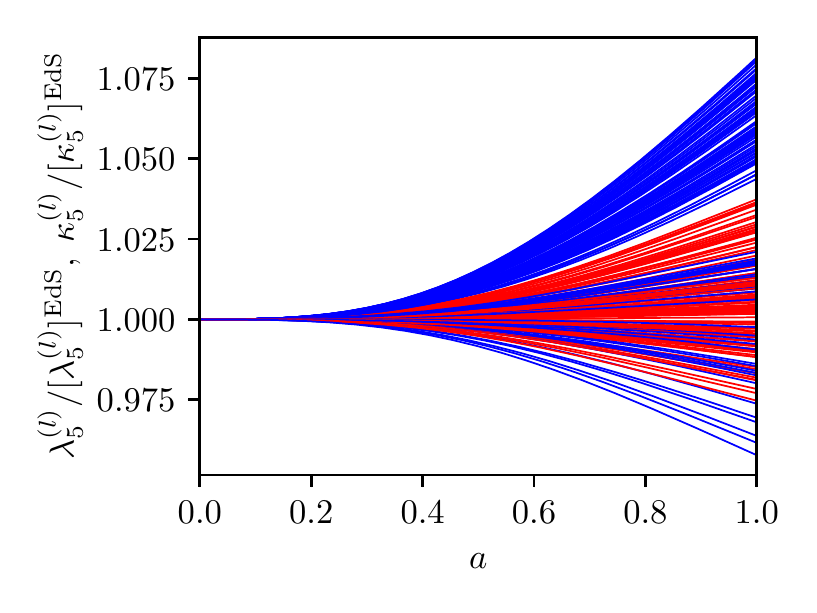}
		\caption{$n = 5$}
	\end{subfigure}
	\caption{\protect{\label{fig:all}}Normalised solutions to Eqs. (\ref{lambda}) and (\ref{kappa}) up to fifth perturbative order in a $\Lambda$CDM universe, calculated using the Chebyshev spectral method with $N = 6$. The functions are shown divided by their EdS counterparts (which are constant in time).}
\end{figure}

Figure~\ref{fig:Comparison1} shows a direct comparison between results for $\lambda_3^{(1)}$ as calculated by direct numerical integration of the differential equations and by the CSM with $N \in [2,4,6]$ components. Results obtained using the EdS approximation are also shown for illustrative purposes. Both the residues are shown (top) as well as the absolute value of the relative error (bottom), displaying several clear facts. First, the CSM successfully manages to reproduce the dynamics of this third-order coefficient, with an expansion truncated at $N=2$ already achieving an accuracy of greater than 0.03\%. Furthermore, we find the expected result that increasing $N$ yields more accurate results. This suggests that the user is effectively able to control the output accuracy of results, though with slight sacrifices in efficiency (as discussed in Section \ref{sec:spectraltest}). Finally, we also find that the relative accuracy of the CSM tends to improve at late times ($a\rightarrow1$) away from zero crossings for all component numbers and at all orders.

\begin{figure}[ht!]
	\centering
	\begin{subfigure}[b]{0.9\textwidth}
		\centering
        \includegraphics{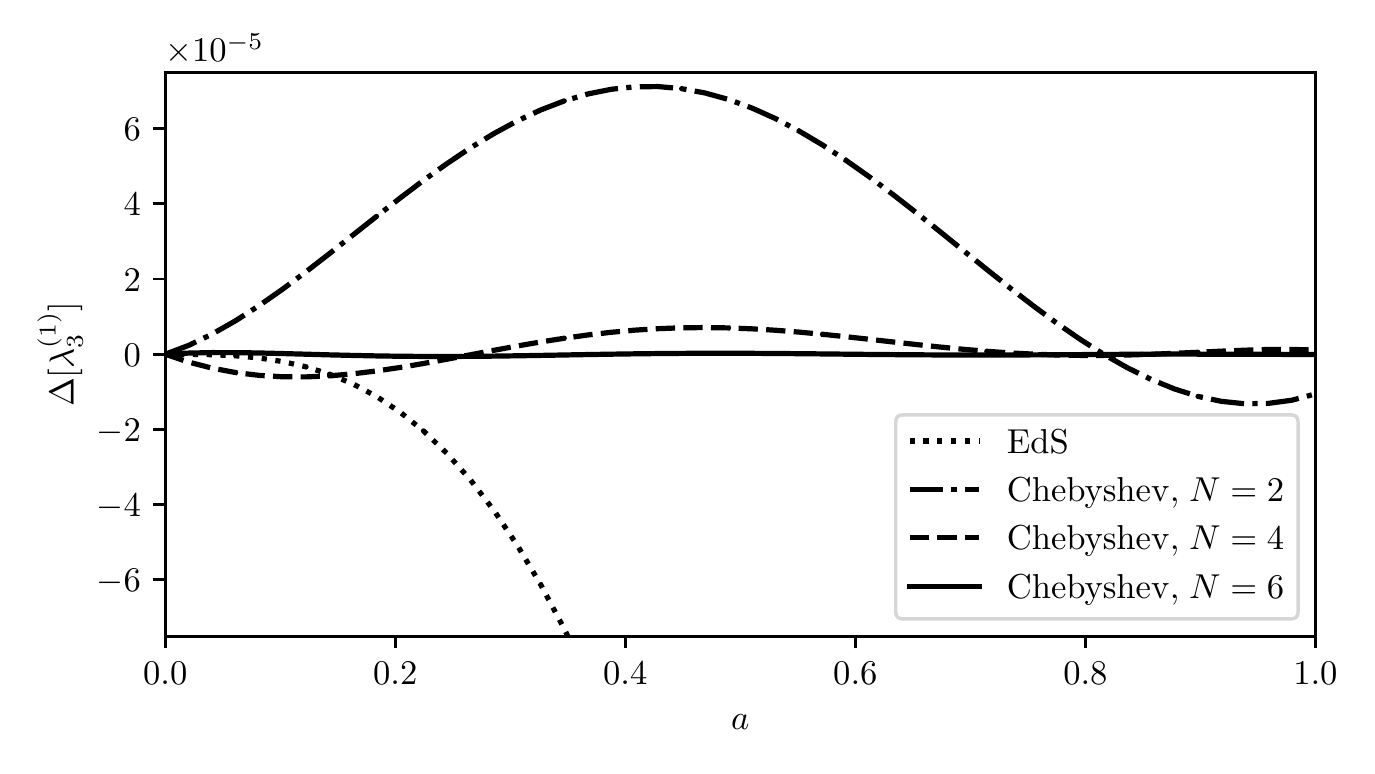}
	\end{subfigure}
	
	\vskip\baselineskip
	\begin{subfigure}[b]{0.9\textwidth}
		\centering 
        \includegraphics{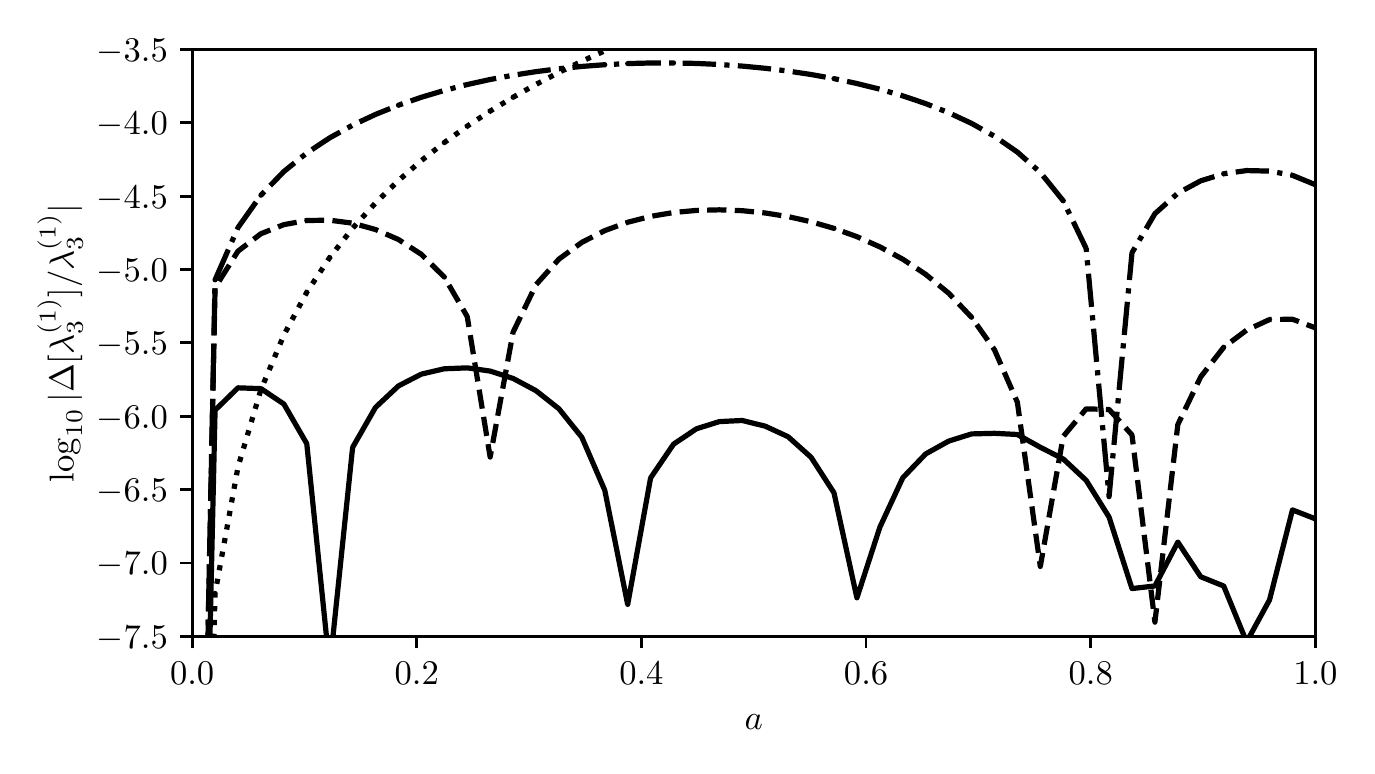}
	\end{subfigure}
	\caption{Comparison of approximate results for $\lambda_3^{(1)}$ from the Chebyshev spectral method and direct numerical integration. Differences with respect to the direct solution are shown in the top panel keeping different numbers of terms in the expansions in Chebyshev polynomials, while the (absolute values) of the fractional differences are shown in the bottom panel. The Chebyshev spectral method is very accurate for all $a$ even with few terms in the expansion. On the other hand, the EdS approximation  (dotted lines) diverges rapidly away from the true solution.}
    \label{fig:Comparison1}
\end{figure}

Next, Fig. \ref{fig:WmDependence} shows the normalised dynamical dependence of $\lambda_5^{(1)}(a)$ as a one-parameter-family of $\Omega_{m_0}$, calculated by the CSM with $N = 5$. Here, examples of dark-energy dominated (red), $\Lambda$CDM (black) and matter-dominated (blue) cosmologies are shown, along with the shaded full parameter space. Here, it is found that $\lambda_5^{(1)}$ shows a non-trivial $\Omega_{m_0}$ dependence, further implying the importance of fully taking these dynamics into account. Finally, the CSM successfully reproduces the expected EdS result of constant coefficients as $\Omega_{m_0} \rightarrow 1$. 

\begin{figure}[ht!]
    \centering
    \includegraphics{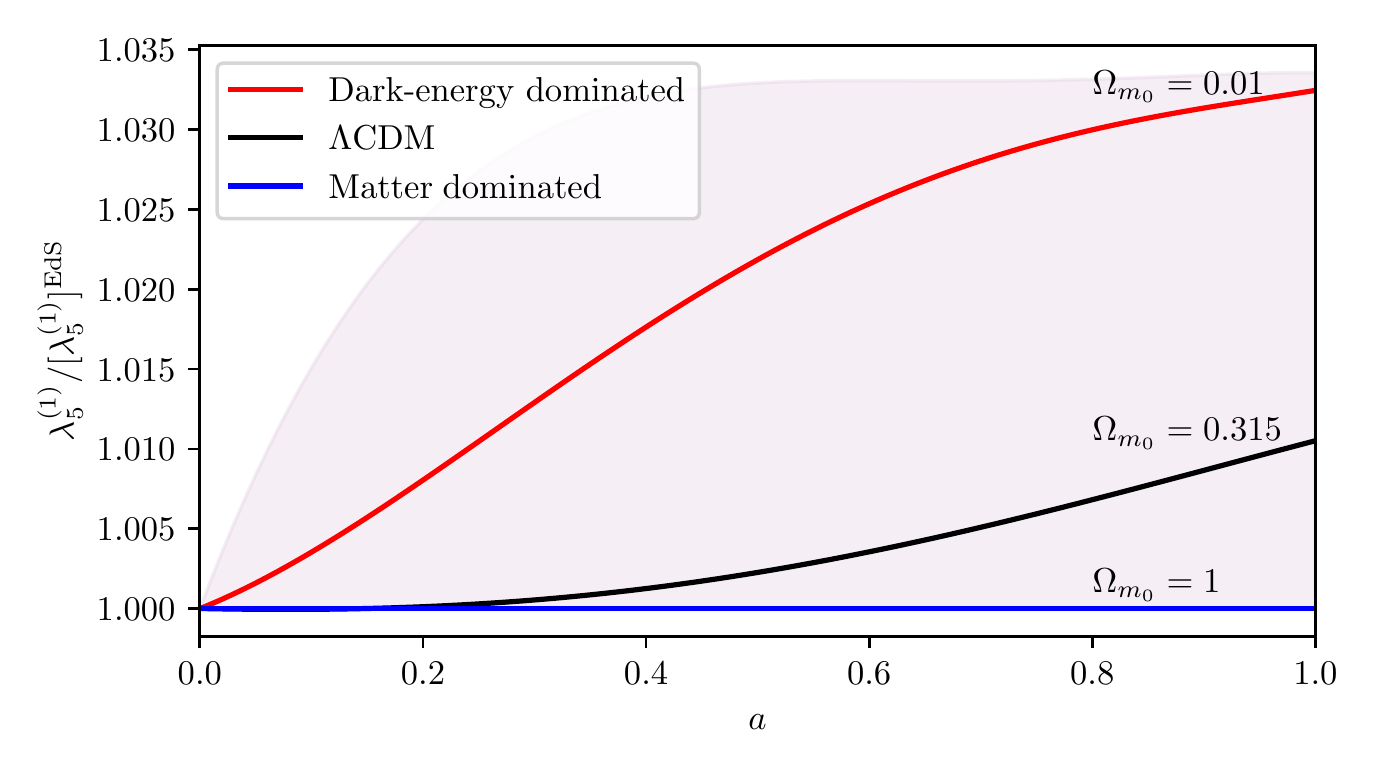}
    \caption{Dynamical results for $\lambda_5^{(1)}$ calculated using the CSM as a one-parameter-family of $\Omega_{m_0}$. Specifically, matter-dominated (blue), $\Lambda$CDM (black) and dark-energy-dominated (red) results are highlighted, with matter density parameters labelled. The boundary of the shaded region is for $\Omega_{m_0}\rightarrow 0$. }
    \label{fig:WmDependence}
\end{figure}

The final comparison we make is to an alternative solution to the problem considered in this paper, which we shall introduce here following results in \citep{Fasiello:2022lff}.
The key observation there is that the parametric dependence of the equations of motion~\eqref{alpha} and~\eqref{beta} on $\Omega_{m_0}$ can be absorbed into a new independent variable $q$, where
\begin{equation}
 q \equiv \left(\frac{1 - \Omega_{m_0}}{\Omega_{m_0}}\right)a^3 .
\end{equation}
This follows by noting that $D_+(a;\Omega_{m_0})$ may be written as
\begin{equation}
D_+(a;\Omega_{m_0}) = \left(\frac{\Omega_{m_0}}{1-\Omega_{m_0}}\right)^{1/3} \hat{D}_+(q) ,
\end{equation}
where the rescaled growth function $\hat{D}_+$ depends only on $q$. It follows that $f_\pm(a;\Omega_{m_0})$ also depend only on $q$ and, since $d \ln D_+ = d \ln \hat{D}_+$, so do the solutions $U_{\alpha/\beta}$ and $W_{\alpha/\beta}$ of Eqs~\eqref{alpha} and~\eqref{beta}. Expanding $f_\pm(q)$ as power series in $q$, or, equivalently, in $\zeta \equiv \hat{D}_+^3$ on noting that around $\Omega_{m_0} = 1$ (i.e., $q=0$) $\hat{D}_+(q) = q^{1/3} \left[1+\mathcal{O}(q)\right]$, we have

\begin{equation}
\frac{f_-}{f_+^2}(a;\;\Omega_{m_0}) = -\frac{3}{2} + \sum_{i = 1}^{\infty} c_i \zeta^i .
\end{equation}

Truncating this expansion at low order works well, with~\cite{Fasiello:2022lff} suggesting dropping $c_4$ and higher. Expanding the $\lambda_n^{(l)}$ and $\kappa_n^{(l)}$ similarly, 
\eq{
\lambda_n^{(l)}(a;\Omega_{m_0}) \approx \sum_{i = 0}^3 [g_n^{l}]_i c_i\zeta^i,\qquad
\kappa_n^{(l)}(a;\Omega_{m_0}) \approx \sum_{i = 0}^3 [h_n^{l}]_i c_i\zeta^i ,
\label{eq:perturb}
}
the coefficients may be determined analytically from the equations of motion.

Figure~\ref{fig:perturbative} compares this truncated power-series expansion in $\zeta$ and the CSM, providing plots of the relative errors for all $\lambda_n^{(l)}$ and $\kappa_n^{(l)}$ for perturbative orders $n\in[2,3,4]$ compared to the direct numerical solutions. Though all $\zeta$-expansion (blue) and CSM with $N = 4$ (red) and $N = 6$ (orange) curves are given, the averages have been extracted and shown in bold for clarity. Here, it is found that although both methods are very successful at reproducing all solutions, we find that the CSM consistently produces greater accuracy as $a \rightarrow1$, making it more useful for computation of late-time observations. The CSM is also helped by greater efficiency in calculating components and significantly greater tunability in both accuracy and parameter space, as discussed later in Sec.~\ref{sec:conclusion}. 

\begin{figure}[ht!]
	\centering
	\begin{subfigure}[b]{0.475\textwidth}
		\centering
        \includegraphics{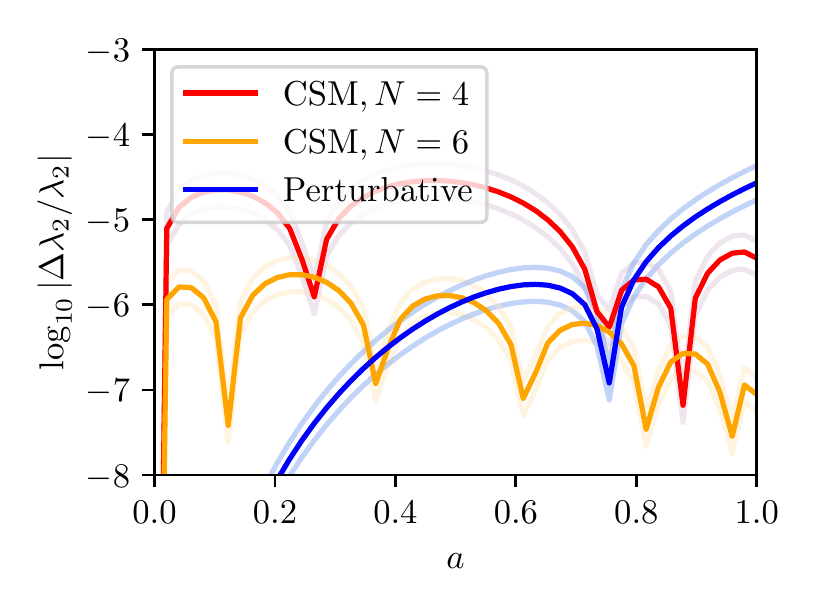} 
	\end{subfigure}
	\hfill
	\begin{subfigure}[b]{0.475\textwidth}
		\centering 
        \includegraphics{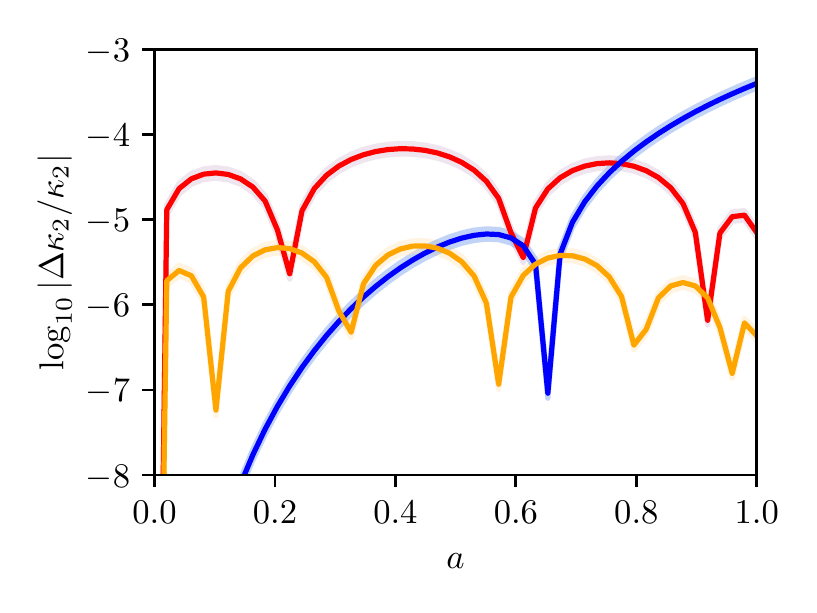}
	\end{subfigure}
	\vskip\baselineskip
	\begin{subfigure}[b]{0.475\textwidth}
		\centering 
        \includegraphics{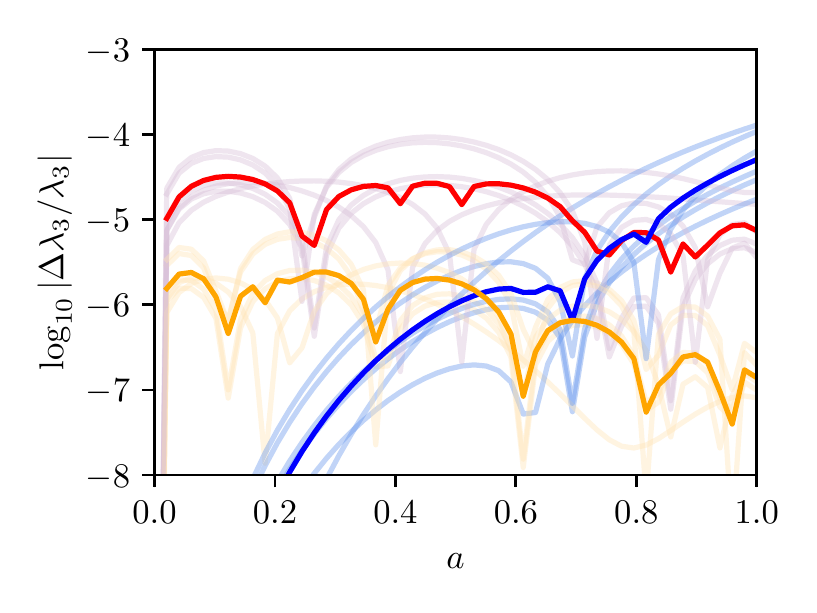}
	\end{subfigure}
	\hfill
	\begin{subfigure}[b]{0.475\textwidth}   
		\centering 
        \includegraphics{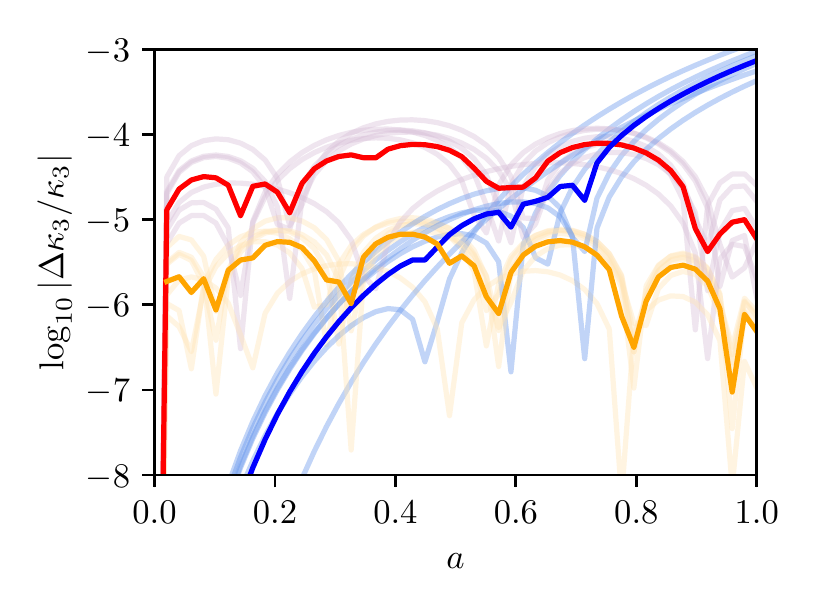}
	\end{subfigure}
    \vskip\baselineskip
	\begin{subfigure}[b]{0.475\textwidth}
		\centering 
        \includegraphics{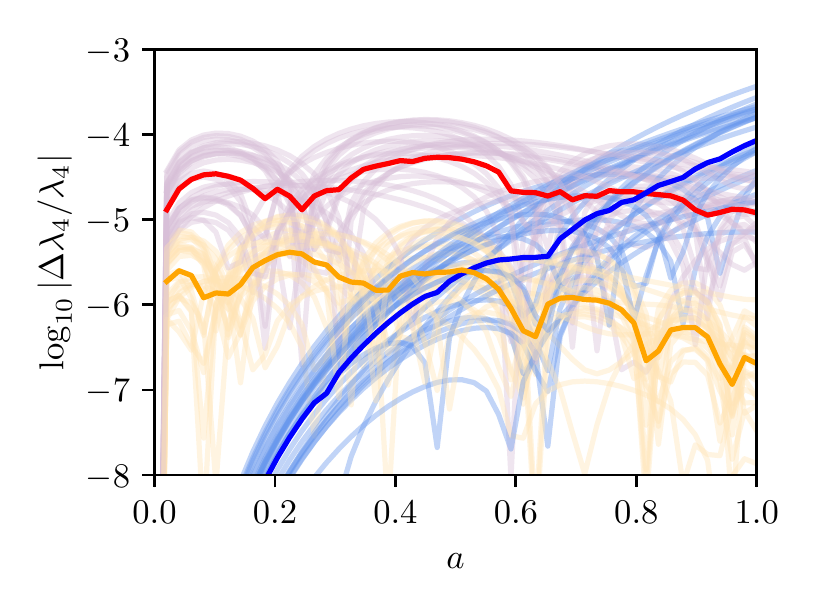}
	\end{subfigure}
	\hfill
	\begin{subfigure}[b]{0.475\textwidth}   
		\centering 
        \includegraphics{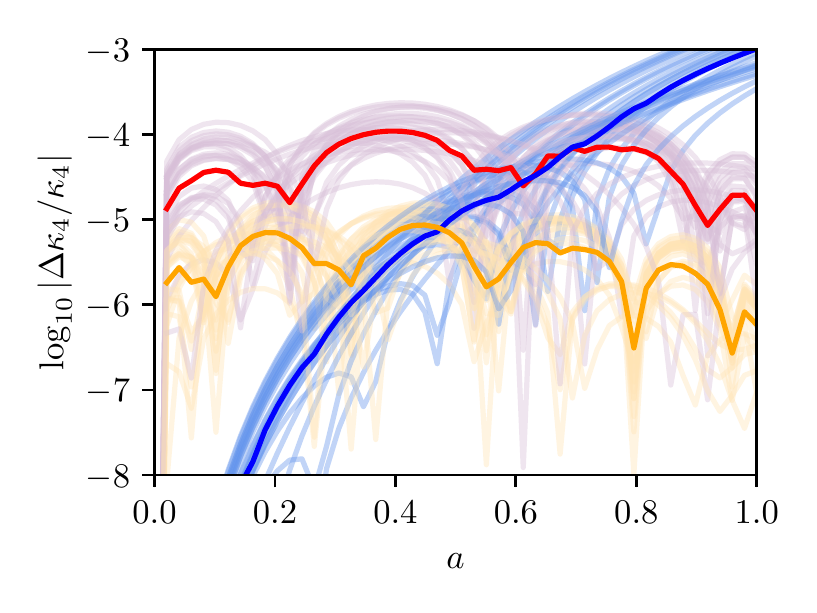}
	\end{subfigure}
	\caption{Comparisons between the relative errors in solutions found for the dynamical coefficients $\lambda_n^{(l)}$ and $\kappa_n^{(l)}$ for the perturbative orders $n=2,3,4$ in the $\Lambda$CDM universe. Results are shown for the truncated power-series expansion in $\zeta$ given by Eq.~\eqref{eq:perturb} (blue; labelled Perturbative) and the CSM with $N = 4$ (red) and $N = 6$ (orange). Curves are shown for all $l$ modes, with averages given in bold.}
    \label{fig:perturbative}
\end{figure}

Finally, Fig. \ref{fig:PSE} shows the accuracy of the CSM as applied to calculating the present-day one- (left) and two-loop (right) matter--matter (black), matter--velocity (red) and velocity--velocity (blue) power spectra compared to numerically evaluated spectra. These were produced by combining dynamical coefficients produced by the CSM with momentum kernels calculated as described in \cite{Fasiello:2022lff}, omitted here for brevity. Here, we show both the loop-order contributions to the $\Lambda$CDM power spectrum (top) and absolute-value relative error plots (${\rm Er}(x) = \log_{10}(|\Delta x|/x)$; bottom) for the overall $\Lambda$CDM power spectrum up to each loop order using $N = 2$ (dotted), $N = 4$ (dot-dashed) and $N = 6$ (solid) Chebyshev components. Clearly, in the case of the one-loop power spectrum, using $N = 2$ components is sufficient to produce results to sub-percent-level accuracy. In the case of the two-loop power spectrum contributions, it was found that such gains were dwarfed by numerical uncertainties produced by the numerical integration of the momentum kernels. In principle, however, this demonstrates the effectiveness of the CSM in predicting actual observables.

\begin{figure}[t!]
\centering
	\begin{subfigure}[b]{0.475\textwidth}
		\centering
        \includegraphics{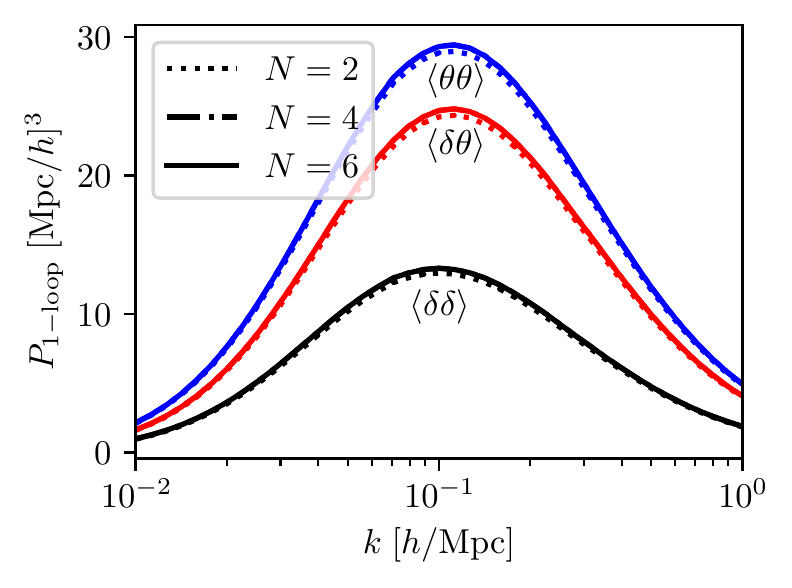} 
	\end{subfigure}
	\hfill
	\begin{subfigure}[b]{0.475\textwidth}
		\centering 
        \includegraphics{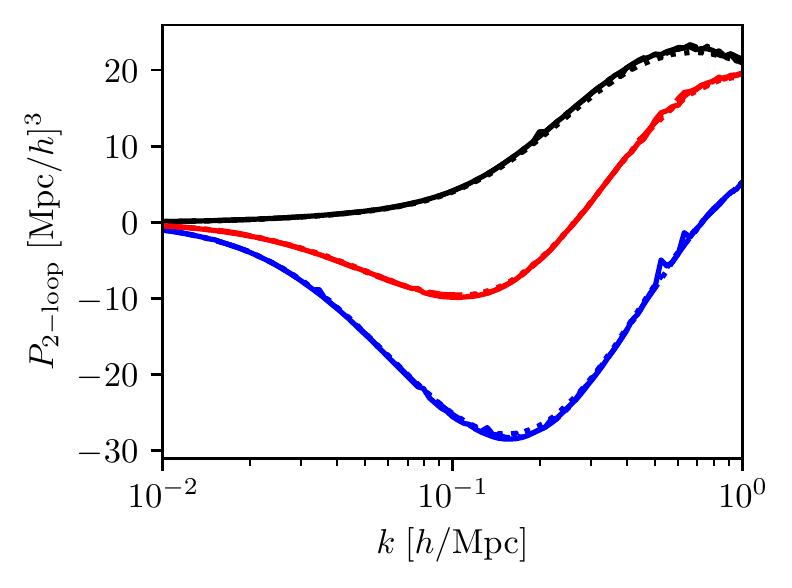}
	\end{subfigure}
	\vskip\baselineskip
	\begin{subfigure}[b]{0.475\textwidth}
		\centering 
        \includegraphics{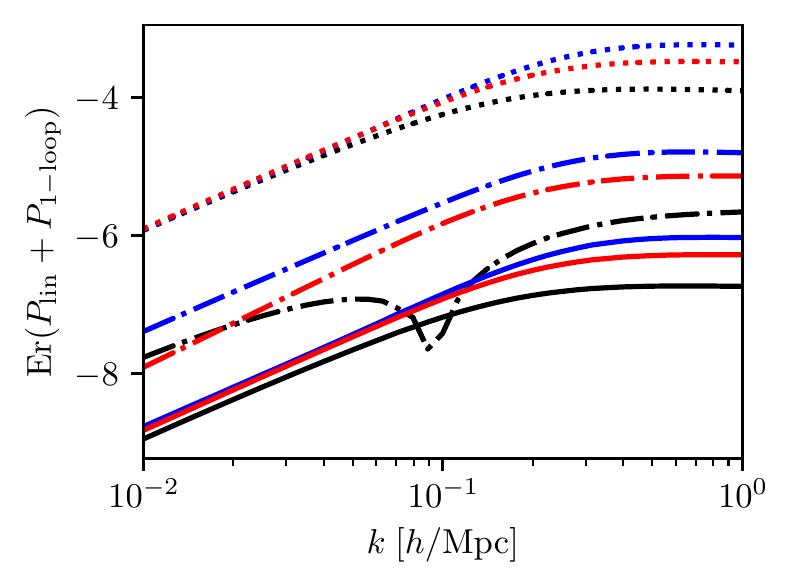}
	\end{subfigure}
	\hfill
	\begin{subfigure}[b]{0.475\textwidth}   
		\centering 
        \includegraphics{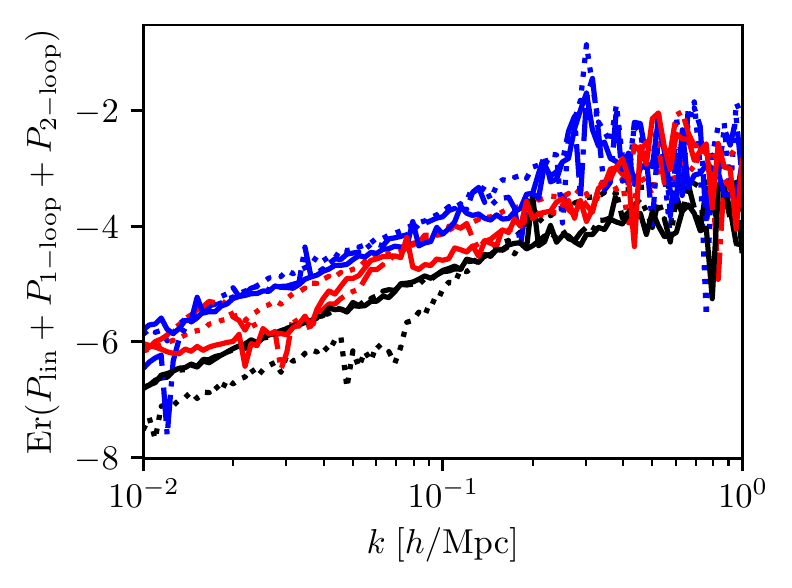}
	\end{subfigure}

\caption{Present-day one-loop (left) and two-loop (right) contributions to the matter--matter (black), matter--velocity (red) and velocity--velocity (blue) power spectra, computed using the CSM with $N = 2$ (dotted), $N = 4$ (dot-dashed) and $N = 6$ (solid) Chebyshev polynomials. We show the loop-order contributions to the $\Lambda$CDM power spectrum (top) and relative errors in the $\Lambda$CDM power spectra as compared to the numerical values (${\rm Er}(x) = \log_{10}(|\Delta x|/x)$; bottom) for each case.  Errors in the two-loop contributions are dominated by numerical uncertainties in calculations of the momentum kernels.}
\label{fig:PSE}
\end{figure}

\subsection{Benchmarking the Code}
\label{sec:benchmarking}

In Fig.~\ref{fig:timingcomp1} we compare the average computational time required to solve Eqs.~(\ref{lambda}) and (\ref{kappa}) with direct numerical integration (as in Sec.~\ref{sec:numerical}) and our implementation of the CSM. Specifically, we show the time taken to calculate all dynamical components for a given perturbative order $n$ for the Planck best-fit $\Lambda$CDM universe. For the direct method (red) we use 50 subdivisions of the range $a\in [0,1]$, while for the CSM (blue) we use a variety of components in the Chebyshev expansion ($N \in [3,4,10]$).
Of these, $N = 4$ would be sufficiently accurate for most uses (and is thus discussed below), while $N = 10$ is far more accurate than necessary for the time taken. We find that the terms needed for the one-loop power spectrum ($n = 2$ and $3$) are calculated two orders of magnitude faster than by direct numerical integration; those for the two-loop power spectrum ($n = 2,3,4,5$) are calculated more than four orders of magnitude faster. We also comment that the CSM code used here is inherently iterative, therefore returning the results for all previous-order coefficients while calculating the dynamics of a target $n$. This further exemplifies the efficiency of this method, particularly within the context of its flexibility. Finally, due to this iterative nature, our implementation of the CSM is able to calculate the dynamical coefficients to any perturbative order. This implies that this code will remain useful as momentum operators at increasingly greater orders are computed in the future.

\begin{figure}[ht!]
    \centering
    \includegraphics{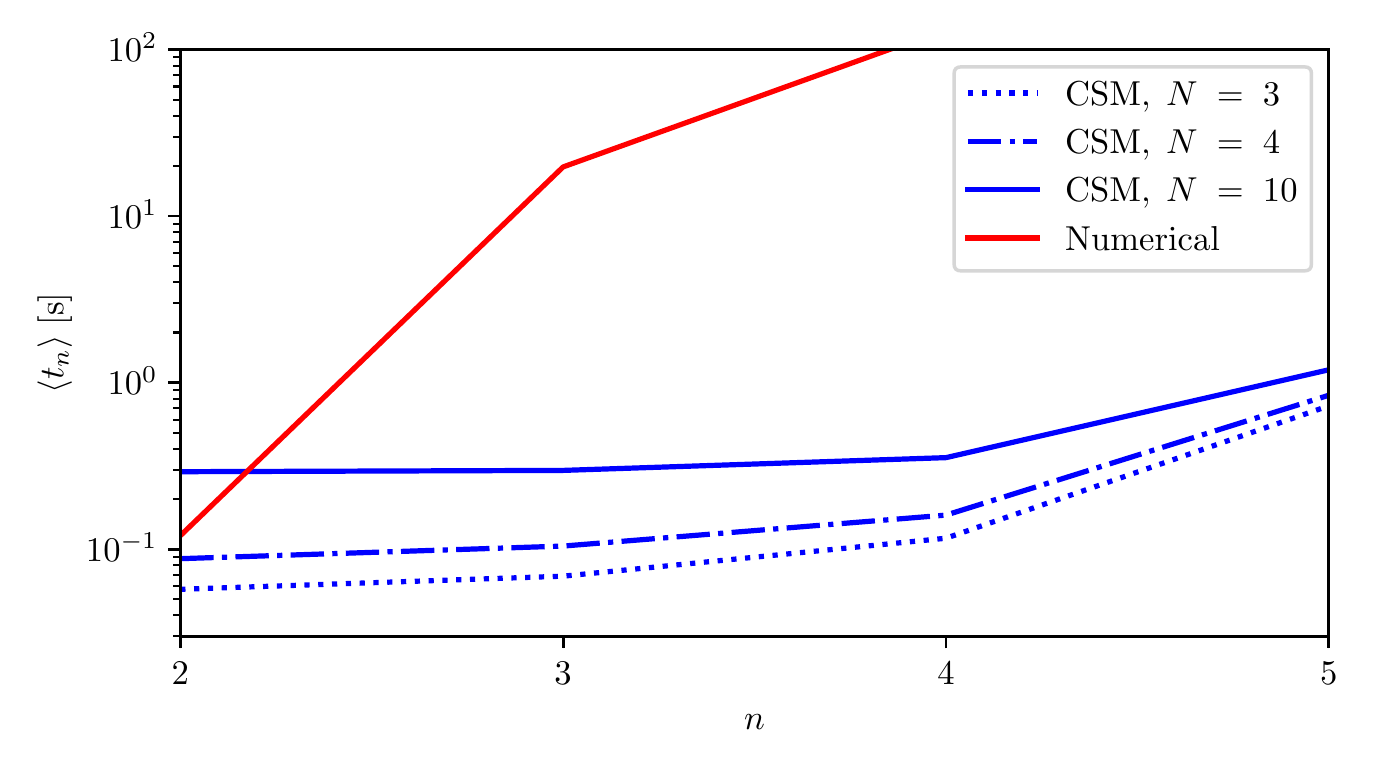}
    \caption{\protect{\label{fig:timingcomp1}}Comparison of the time taken to compute all $\lambda_n^{(l)}$ and $\kappa_n^{(l)}$ for varying perturbative order $n$ in the Planck best-fit $\Lambda$CDM universe using both direct numerical integration (red) and Chebyshev spectral methods (blue) with varying degrees of accuracy. Direct numerical solutions for higher $n$ are not included as their computation is prohibitively slow.}
\end{figure}

\section{Discussion and Conclusions} 
\label{sec:conclusion}

The results in Section \ref{sec:results}, including, most notably, Figs.~\ref{fig:Comparison1}, \ref{fig:perturbative} and \ref{fig:PSE} demonstrate the accuracy and effectiveness of the Chebyshev spectral method for solving the full dynamical evolution of the dark matter density and velocity fluctuation fields in a $\Lambda$CDM universe. This is coupled with the fact that the CSM is, in practice, at least an order of magnitude faster than direct numerical ODE solution methods. The true utility of this method is felt, however, at higher perturbative orders $n$, with Fig.~\ref{fig:timingcomp1} demonstrating how much faster the CSM is at such orders.

From a practical standpoint, the code implementation produced as part of this work has the following features. First, the method can be used with any number of Chebyshev components $N$. Doing so increases the accuracy of the resultant solution somewhat but incurs an extra computational time cost, as seen in Fig.~\ref{fig:timingcomp1}. As such, the user is able to make this decision actively and tailor the method to their particular situation and use case. Next, it is found that the magnitude of the calculated components drops almost exponentially with $N$, further showing the futility of finding these high-order components. As a result, in testing it was found that using $N = 4$ components is more than suitable, with $N = 2$ yielding sufficient results in most cases, as can be seen explicitly in Figs.~\ref{fig:Comparison1} and \ref{fig:PSE}. 

Furthermore, the code has been generalised to work for arbitrary $\Omega_{m_0}$, extending the parameter space of this method to encompass $\Lambda$CDM. The effect of this has been explored in Fig.~\ref{fig:WmDependence}. Along these lines, we have also compared the effectiveness of the CSM to an alternative perturbative method introduced in~\citep{Fasiello:2022lff}, which is also valid for $\Lambda$CDM only. 
It was found that the CSM was able to achieve a similar or better degree of accuracy, particularly for late times. The CSM also grants the user a choice between accuracy and efficiency, which is important when calculating model predictions for cosmological likelihood analyses.
Furthermore, while the perturbative method is on the surface faster to evaluate for a given $\Omega_{m_0}$, the CSM allows the user to generate dynamical-coefficient data to any perturbative order efficiently. Next, the framework discussed in Section \ref{sec:LCDM_dynamics} can be extended to
a more general set of cosmologies, including clustered quintessence~\citep{Sefusatti:2011cm, Fasiello:2019}. On the other hand, it is thought that the perturbative method outlined above in Eq.~\eqref{eq:perturb} will only work in the case of $\Lambda$CDM. Indeed, such an extension of the CSM has already been done to first-loop-order for the power spectrum, with work underway to expand this further.\footnote{This refers to work completed in a masters thesis by one of the authors, along with an upcoming paper.}

Another interesting direction to explore would be to consider the implications of other models of dark matter and energy. These might include decaying dark matter~\citep{Bell:2011} or a dynamical dark energy equation of state \citep{Chevallier:2001, Linder:2003}. Finally, it would be particularly interesting to explore the impact of various dark energy effective field theories (EFTs), of which clustered quintessence is an example \citep{Ferreira:1997,Zlatev:1999, Frusciante:2020, Aviles2:2018, Sefusatti:2011cm, Fasiello:2019, Wang:1998}. However, the most natural extension would be to implement similar methods to study other observables to the same loop order, such as the bispectrum.

To conclude, we have shown how the CSM can optimize the problem of solving the full, perturbative dynamics of perturbations in the $\Lambda$CDM Universe, reducing computation time by several orders of magnitude, with a view of being potentially extended to a more general set of cosmologies. Our implementation of the CSM as a Python library is presented alongside this paper (\url{https://github.com/Chousti/CSMethod.git}). It should be straightforward to integrate this library into galaxy-clustering likelihood frameworks, allowing accurate predictions of observables in the perturbative regime fully accounting for the $\Lambda$CDM dynamics.

\begin{acknowledgments}
We thank Tomohiro Fujita and Matteo Fasiello for useful discussions. 
N.C.\ and Z.V.\ acknowledge the support of the Kavli Foundation. N.C. acknowledges support from the Science and Technology Facilities Council (STFC) for a Ph.D. studentship. A.C.\ acknowledges support from the STFC (grant numbers ST/N000927/1 and ST/S000623/1).
\end{acknowledgments}

\appendix

\section{Auxiliary Functions}
\label{sec:appendixI}

Here we review all auxiliary functions necessary to calculate Eqs. (\ref{lambda}) and (\ref{kappa}). These are derived in \citep{Fasiello:2022lff} and represent a recursive method to establish the maximum number of dynamical coefficients. In principle, this number can be reduced by utilising physical constraints such as conservation of mass and momentum or the equivalence principle --- though these are not discussed here.
Specifically, we have the numbering function,
\eeq{
N(n) = \frac{1}{2}\df^K_{\frac{n}{2},\lfloor\frac{n}{2}\rfloor}N(n/2)[1 + 3N(n/2)]  + 3\sum_{m = 1}^{\lfloor (n - 1)/2\rfloor}N(m)N(n - m),
}

and bijective maps,
\eq{
	\phi_1(n,i,j) &= N(n/2)(i - j) + i,\\
	\phi_2(n,i,j) &= [N(n/2)]^2 - \frac{1}{2}i(i-1) + \phi_1(n,i,j),\\
	\phi_3(n,m,i,j) &= \frac{1}{2}\df^K_{\frac{n}{2},\lfloor\frac{n}{2}\rfloor} N(n/2)[1+3N(n/2)] + \sum_{k = 1}^{m-1}N(k)N(n - k) + (i-1)N(n-m) + j,\\
	\phi_4(n,m,i,j) &= \sum_{k = 1}^{\lfloor (n-1)/2\rfloor}N(k)N(n-k) + \phi_3(n,m,i,j),\\
	\phi_5(n,m,i,j) &= 2\sum_{k = 1}^{\lfloor (n-1)/2\rfloor}N(k)N(n-k) + \phi_3(n,m,i,j).
}
The CSM makes use of these by looping over all maps while evaluating Eqs. (\ref{alpha}) and (\ref{beta}), hence algorithmically calculating all necessary functions.


\bibliography{Reference}

\end{document}